\documentclass[12pt]{article}
\usepackage{amsmath,amssymb,amsthm}
\usepackage{latexsym,graphicx,color,subfigure,mathrsfs}

\def\6#1{{\underline{#1}}}
\def\m6#1{{\underline{#1}\,}}

\newdimen\Tdim
\def\ispan{{\setbox0=\hbox{i}%
\Tdim\ht0\advance\Tdim\dp0\rule[-\dp0]{0pt}{\Tdim}}}
\def\jspan{{\setbox0=\hbox{j}%
\Tdim\ht0\advance\Tdim\dp0\rule[-\dp0]{0pt}{\Tdim}}}
\def\Tspan#1{{\setbox0=\hbox{#1}%
\Tdim\ht0\advance\Tdim\dp0\advance\Tdim.55ex\rule[-\dp0]{0pt}{\Tdim}\box0}}

\def\be{\begin{eqnarray}}
\def\ben{\begin{eqnarray*}}
\def\ee{\end{eqnarray}}
\def\een{\end{eqnarray*}}
\def\Tr{{\rm Tr}}
\def\tr{{\rm tr}}

\def\p{\partial}

\def\D{\mathcal{D}}

\def\=:{=\hspace{-.7em}\raisebox{1.1ex}{.}\hspace{.1em}\raisebox{-0.2ex}{.} }

\newcommand{\NF}{N_{\rm f}}

\newcommand {\beq}{\begin{eqnarray}}
\newcommand {\eeq}{\end{eqnarray}}
\newcommand {\non}{\nonumber\\}

\def\diag{{\rm diag}}

\theoremstyle{definition}

\makeatletter
\@addtoreset{equation}{section}

\renewcommand{\thefootnote}{\fnsymbol{footnote}}
\makeatother

\makeatletter
\newcommand{\thetablename}{Table}
\def\fnum@table{\thetablename\ \thetable}
\makeatother

\begin{document}

\thispagestyle{empty}
\begin{flushright}
RIKEN-MP-4\\
IFUP-TH/2010-23
\end{flushright}
\vspace{3mm}

\begin{center}
{\Large \bf Zero-modes of Non-Abelian Solitons \\
in Three Dimensional Gauge Theories}  
\\[15mm]
{Minoru~{\sc Eto}}$^{1}$\footnote{\it e-mail address:
meto(at)riken.jp},
{Sven Bjarke~{\sc Gudnason}}$^{2,3}$\footnote{\it e-mail address:
gudnason(at)df.unipi.it}
\vskip 6 mm

\bigskip\bigskip
{\it
$^1$~Mathematical Physics Lab., RIKEN Nishina Center, Saitama 351-0198, Japan\\
$^2$ Department of Physics, Enrico Fermi, University of Pisa,
Largo Bruno Pontecorvo, 3, Ed. C, 56127 Pisa, Italy\\
$^3$ INFN, Sezione di Pisa,
Largo Bruno Pontecorvo, 3, Ed. C, 56127 Pisa, Italy
}

\bigskip

\bigskip

{\bf Abstract}\\[5mm]
{\parbox{13cm}{\hspace{5mm}
\small
We study non-Abelian solitons of the Bogomol'nyi type in
$\mathcal{N}=2$ ($d=2+1$) supersymmetric Chern-Simons (CS) and Yang-Mills (YM) theory
with a generic gauge group.
In CS theory, we find topological, non-topological and
semi-local (non-)topological vortices of non-Abelian kinds
in unbroken, broken and partially broken vacua. 
We calculate the number of zero-modes using an index theorem and then
we apply the moduli matrix formalism to realize the moduli
parameters. 
For the topological solitons we exhaust all the moduli while
we study several examples of the non-topological and semi-local
solitons. 
We find that the zero-modes of the topological solitons are governed 
by the moduli matrix $H_0$ only and those of the non-topological
solitons are governed by both $H_0$ and the gauge invariant field
$\Omega$. 
We prove local uniqueness of the master equation in the YM case and
finally, compare all results between the CS and YM theories. 

}}
\end{center}
\newpage
\pagenumbering{arabic}
\setcounter{page}{1}
\setcounter{footnote}{0}
\renewcommand{\thefootnote}{\arabic{footnote}}

\tableofcontents

\section{Introduction}

Chern-Simons (CS) theory is an alternative gauge theory in odd
space-time dimensions (we only consider $d=2+1$ here), which is very
different from Yang-Mills (YM) and Maxwell gauge theories. 
The CS gauge fields in pure CS theory are
non-dynamical, however, they become very interesting upon coupling
with other fields.
For instance, when considering a CS kinetic term together with a YM
kinetic term, the gauge bosons acquire a topological mass in the
absence of the Higgs mechanism \cite{Deser:1981wh}. 
The CS gauge theories are interesting both due to their theoretical
beauty and also for their experimental applications, such as the
fractional quantum Hall effect \cite{Zhang:1988wy}. 
The CS term can be induced by quantum radiative corrections, even if
the original Lagrangian does not include a bare CS term
\cite{Redlich:1983kn}, such as the $d=2+1$ dimensional quantum
electrodynamics (QED).

One of the important features of both CS and YM theories coupled with
scalar fields is the existence of solitonic solutions to the classical
equations of motion. Thousands of works are made on the study of these
solitons. 
For the CS solitons, early works with/without the Maxwell term were
studied in Refs.~\cite{deVega:1986eu,Paul:1986ix}.
In the Abelian CS theory coupled with a complex scalar field, solitons
of the self-dual type, viz.~solutions satisfying the Bogomol'nyi
energy bound and in turn requiring a special sixth-order scalar
potential were found in Refs.~\cite{Hong:1990yh,Jackiw:1990aw}. 
The supersymmetric version which corresponds to the self-dual model
was obtained in
Ref.~\cite{Lee:1990it} and the study on zero-modes of the solitons
was developed in Ref.~\cite{Jackiw:1990pr}.
Furthermore, due to the presence of both the CS vacuum and
the Higgs vacuum, domain walls and non-topological vortices were found 
in addition to the familiar topological vortices in the Abelian
CS-Higgs (CSH) model \cite{Jackiw:1990pr}. 
The non-topological soliton is a typical soliton in CS theory which
in fact does not exist in the Maxwell-Higgs (or Yang-Mills-Higgs (YMH))
theories. 
Similar solitons of topological/non-topological kinds were found also
in the Maxwell-CS-Higgs (MCSH) theory
\cite{Lee:1990eq,Lee:1991td,Lee:1992yc}. 
Furthermore, a semi-local extension of this type of solitons, viz.~in
MCSH coupled with two Higgs fields was found both in the
topological as well as the non-topological case in
Ref.~\cite{Khare:1992qr}. 
An extension to the non-Abelian ($SU(N)$) CS theories was also made and
non-topological solitons with a global charge were found in
Ref.~\cite{Lee:1990ep}. There are also lots of interesting works on
the solitons in the non-relativistic CS theories.
Many relevant references on both the relativistic and non-relativistic
models can be found in the excellent reviews
\cite{Dunne:1998qy,Horvathy:2008hd}. 
The mentioned works on the CS solitons were established in the early `90s
and we shall refer to all these as solitons of the Abelian
kind.\footnote{In CS theory as well as YM theory, the vortex in
  $SU(N)$ theories is basically of the Abelian kind due to the first
  homotopy group characterizing the vortices being
  $\pi_1(SU(N)/\mathbb{Z}_N)=\mathbb{Z}_N$ and hence these vortices do
  not carry non-Abelian orientational moduli, as opposed to those we
  discuss next.}

The second era of the CS solitons recently started when a new type of
the toplogical CS soliton of the non-Abelian (NA) kind\footnote{The
  non-Abelian vortices were first found in the supersymmetric YM
  theories in four dimensions \cite{Hanany:2003hp,Auzzi:2003fs}. There
  are many works on the NA vortices in YM theories and we shall refer
  to the literature for the many results which however are nicely
  summarized in several review articles \cite{review,Eto:2006pg}.} was
found in 
the $U(N)$ CS theory coupled with $N$ Higgs fields in the fundamental
representation \cite{Aldrovandi:2007nb}. 
The NA CS vortex found so far is quite similar to the NA YM vortex
\cite{Hanany:2003hp,Auzzi:2003fs,review,Eto:2006pg}.
It lives in the color-flavor
locked, broken vacuum (i.e.~the Higgs vacuum) and has internal
orientational zero-modes,  $\mathbb{C}P^{N-1}$. 
These internal orientational modes are related to the color and flavor
degrees of freedom, which make the solitons being truly of the NA
kind. 
Soon after its discovery, the NA semi-local vortices were found in the 
$U(N)$ CS theory with $\NF>N$ fundamental Higgs fields
\cite{Lozano:2007yz,Buck:2009pd}. 
The dynamics of the NA vortices was studied in
Ref.~\cite{Collie:2008mx} and the NA vortices with a globally
conserved charge was found in Ref.~\cite{Collie:2008za}. 
The topological solitons of both local, semi-local and fractional kind
in the NA CS theory with an arbitrary gauge group of the form $U(1)$
times a simple group have also been studied very recently
in Refs.~\cite{Gudnason:2009ut,Gudnason:2010yy}. 
Thus, the study on the CS solitons is being revived now.

The NA CS solitons found so far are all topological solitons in the
broken vacuum (the Higgs vacuum). This is because their discovery
\cite{Aldrovandi:2007nb} was inspired by the NA YM vortex which is
indeed a topological soliton living in the broken (Higgs)
vacuum. However, we know that there are also non-topological solitons 
in the Abelian CS models \cite{Jackiw:1990aw}. So it is natural to ask
ourselves if non-topological or other solitons of the NA kind exist or 
not, in NA CS theories. With this question in mind, 
we will be concerned in this paper with vortices of all possible
non-Abelian kinds 
in the NA CS gauge theories with the gauge group $G$ being 
general; $G=U(1)\times G'$ with $G'$ is an arbitrary simple group. 
As if it was not broad enough a spectrum, we will make every statement 
throughout the paper be independent of the gauge group.
Beyond our naive expectation, we will find semi-local NA vortices of
both topological and non-topological types in the models where the
number of flavors is less than or equal to the number of colors.

We will investigate not only single solitons but also multiple
solitons, viz.~we will focus on the zero-modes of the Bogomol'nyi
solutions. 
To this end, we will count the number of zero-modes by using an index
theorem technique. 
Then we will extend the moduli matrix method for explicit realization
of the counted moduli.  
The moduli matrix formalism was first introduced in the studies of
domain wall systems \cite{Isozumi:2004jc,Eto:2005wf} but later
developed to encompass vortex systems \cite{Eto:2005yh}, which has 
proven an invaluable tool in soliton studies~
\cite{Isozumi:2004vg,Eto:2004rz,Eto:2005cp,Eto:2006pg}. 
The moduli matrix formalism has already been introduced to the CS
models in Refs.~\cite{Gudnason:2009ut,Gudnason:2010yy} for the
topological solitons. 
Here we will try to further develop it to cover all the solitonic
solutions in the system, including the non-topological solitons in the
unbroken and partially broken vacua. 
It turns out that the moduli matrix formalism, which has proven itself
incredibly useful for topological solitons in YM theories, provides us 
with a common framework for studying all kinds of solitons not only in
YM but also in CS theories and especially in all vacua of the systems
at hand. 
In order to find similarities and differences among the solitons in YM
and CS theories, we will show all the results parallelly in the two
cases. 
Most of the statements concerning the YM vortex (up to
Sec.~\ref{sec:modulimatrix}) are just a review of the known results,
however written in a more transparent way with respect to the gauge
group. A new result -- and one of the main results of this paper --
for the YM vortex is on the uniqueness problem of the so-called master
equation (Sec.~\ref{sec:uniqueness}).

As we have already mentioned, we will pay special attention to the
moduli space of solutions (or simply moduli space). It is one of the
most important properties of solitons -- of topological or
non-topological kind.  
The moduli space has a certain dimension characterizing the soliton 
and the gauge theory (or string theory configuration) at hand and some
local coordinates on it which can be identified with the bosonic
zero-modes of the system describing the soliton. The number of bosonic
zero-modes gives the aforementioned (complex) dimension of the moduli
space.  
The moduli space per se is a classical statement which eventually, far
in the infrared, will be modified by quantum corrections. However,
this information usually is available only through the knowledge about
the classical moduli space supplemented by calculations of a
corresponding sigma model describing the low-energy effective theory
of the soliton under consideration.
 
A further powerful technique often used in soliton physics is the
moduli space approximation or the so-called geodesic approximation due
to a seminal paper \cite{Manton:1981mp} which was first
applied to monopole scattering and thereafter to a vast number of
other soliton configurations in the literature.
 
Index theorems have proven to be very powerful tools in
physics. Needless to mention is the Witten's index indicating if a
theory at hand can have its supersymmetry broken spontaneously or not 
\cite{Witten:1982df}. There is also the Hopf index theorem in sigma 
models giving the number of vacua in a class theories
\cite{Hori:2000kt}.  
An index closely related to what we are interested in here, is the  
Atiyah-Singer \cite{Atiyah:1968mp} index theorem, which has been used 
to count the physical parameters describing instantons after the
configuration space has been compactified \cite{Bernard:1977nr}. 
Finally, there is a generalization of this index to non-compact
manifolds, namely the Callias index theorem which counts the
zero-modes of a Dirac operator minus the zero-modes of the
corresponding adjoint operator \cite{Callias:1977kg}.
This is a very useful technique in soliton physics and especially in
BPS systems where the fluctuations of the BPS equations can easily be
written as a Dirac operator acting on a (vector) space of
fluctuations. 
To mention a few cases in the literature, this technique
has been applied to monopoles \cite{Weinberg:1979ma} and Abelian-Higgs
vortices \cite{Weinberg:1979er}; to Abelian
CS vortices \cite{Jackiw:1990pr}; to Maxwell-CS
vortices \cite{Lee:1991td}; to non-Abelian $U(N)$ vortices \cite{Hanany:2003hp} 
and to domain walls of
Abelian kind \cite{Lee:2002gv} and non-Abelian kind
\cite{Sakai:2005sp}.

The organization of the paper will be as follows. In
Sec.~\ref{sec:model} we set up our theories,
the YM-Higgs model and the NA CS-Higgs model. 
Here we will review the basic properties including the BPS-equations 
of both theories. In Sec.~\ref{sec:indextheorem}, 
we will derive a generalized formula for
the Callias-type index of a certain class of BPS systems including
both our models under consideration.
In Sec.~\ref{sec:modulimatrix} we will review and develop the moduli
matrix formalism to realize the moduli parameters 
and also explain the new types of solitons, namely the NA
non-topological solitons.  
In Sec.~\ref{sec:uniqueness} we will confront the
long-standing 
problem of the uniqueness of the master equations. 
We find a relation between the variation of the master equations and
the vanishing theorems studied in Sec.~\ref{sec:indextheorem}. We
conclude the paper with a discussion and outlook in
Sec.~\ref{sec:conclusion}. 
Secs.~\ref{sec:index_YM} and \ref{sec:mm_ym} are reviews of known
results, so the reader who is familiar with the NA YM vortices can
skip them.


\section{The model and notation\label{sec:model}}

We begin with the ${\cal N}=2$ supersymmetric Yang-Mills-Chern-Simons (YMCS)
theory coupled with Higgs fields in $d=2+1$ dimensions. 
In order to make the following arguments applicable to a wide class of 
gauge theories, we will not specify the gauge group unless we make
some explicit examples. Indeed, we will take the gauge group to be 
on the form $G=\left(U(1)\times G'\right)/\mathbb{Z}_{n_0}$, where
$G'$ is always a simple group.
When we will make some examples we will use the gauge group $G$ with 
$G'=SU(N),SO(N),USp(N)$. 
In this case we can choose either $N=2M$ or $N=2M+1$ for $SO(N)$
whereas $N=2M$ for $USp(N)$. Here $\mathbb{Z}_{n_0}$ is the center of
$G'$, see Table \ref{tab:n0}. 
\begin{table}[ht]
\begin{center}
\begin{tabular}{c||cccc}
$G'$ & $SU(N)$  & $SO(2M)$ & $USp(2M)$ & $SO(2M+1)$ \\
\hline
\hline
$n_0$ & $N$ & 2 & 2 & 1
\end{tabular}
\caption{\small Various values of the greatest common divisor (gcd) of
  the Abelian charges of all the $G'$ invariants which have a non-zero
  VEV at infinity (in the Higgs phase). } 
\label{tab:n0}
\end{center}
\end{table}

\noindent
We will use the standard convention for the Hermitian generators
\beq
\Tr(t^\alpha t^\beta) = \frac{1}{2}\delta^{\alpha\beta} \ , \quad 
t^0 = \frac{{\bf 1}_N}{\sqrt{2N}} \ ,
\label{eq:liebasis}
\eeq
for $\alpha,\beta = 0,1,2,\ldots,\dim(G')$, while the index of $G'$
will be denoted by $a,b=1,2,\cdots,\dim(G')$. 

The $G'$ vector multiplet contains the gauge fields and a real adjoint
scalar field $\{A_\mu^a,\phi^a\}$. The $U(1)$ vector multiplet also 
has the corresponding gauge fields as well as a real singlet scalar
field $\{A_\mu^0,\phi^0\}$. 
We consider $\NF$ Higgs fields $H_r^A$ 
($r=1,2,\ldots,N;\ A=1,2,\ldots,\NF$) 
in the fundamental representation of $G'$ with uniform $U(1)$ charge. 
In the following we will use a matrix notation where $H$ is an 
$N \times \NF$ dimensional matrix, so that the gauge symmetry acts
from the left-hand side and the flavor symmetry acts from the
right-hand side. 
We will use the following compact notation for the construction of
$\mathfrak{g}$ and $\mathfrak{g}'$ algebra valued fields,
respectively, as follows 
\beq
\phi = \sum_{\alpha=0}^{\dim G'} \phi^\alpha t^\alpha \ ,\qquad
\hat{\phi} = \sum_{a=1}^{\dim G'} \phi^a t^a \ .
\eeq

We are now ready to write down the Lagrangian. Since all the fermions
do not play an important role in the following argument for the 
solitons under consideration, we show only the bosonic part 
\begin{align}
\mathcal{L}_{\rm YMCSH} &=
-\frac{1}{4g^2}\left(F_{\mu\nu}^a\right)^2
+\frac{1}{2g^2}\left(\D_\mu\phi^a\right)^2
-\frac{\mu}{8\pi}\epsilon^{\mu\nu\rho}
  \left(A_{\mu}^a\partial_\nu A_\rho^a  
  -\frac{1}{3}f^{abc}A_{\mu}^a A_{\nu}^b A_{\rho}^c\right)\non
&\phantom{=\ }
-\frac{1}{4e^2}\left(F_{\mu\nu}^0\right)^2
+\frac{1}{2e^2}\left(\partial_\mu\phi^0\right)^2
-\frac{\kappa}{8\pi}
  \epsilon^{\mu\nu\rho} A_{\mu}^0\partial_\nu A_\rho^0 \non
&\phantom{=\ }
+\Tr\left[\D_\mu H\left(\D^\mu H\right)^\dag\right] -V_{\rm YMCSH} \ ,\\
V_{\rm YMCSH} &=
\frac{g^2}{2}\left\{\Tr\left[\left(HH^\dag -\frac{\mu}{2\pi}\phi\right)t^a\right]\right\}^2
+\frac{e^2}{2}\left\{\Tr\left[\left(HH^\dag -\frac{\kappa}{2\pi}\phi - \frac{\xi}{N} {\bf 1}_N\right)t^0\right]
\right\}^2 \ . \label{eq:LYMCSH}
\end{align}
Our conventions are
\beq
F_{\mu\nu} = \p_\mu A_\nu - \p_\nu A_\mu + i\left[A_\mu,A_\nu\right]
\ , \quad
\D_\mu H = \left(\p_\mu + i A_\mu\right)H \ , \quad
\D_\mu\phi = \p_\mu\phi + i \left[A_\mu,\phi\right] \ .
\eeq
The first line is the kinetic term of the $G'$ vector multiplet and
the non-Abelian CS term with the gauge coupling constant $g$
and the CS coupling constant $\mu$ which has to be an
integer in the quantum theory in order to preserve gauge invariance
(up to large gauge transformations). 
The second line is the contribution from the Abelian vector multiplet
with $e$ being the gauge coupling constant and $\kappa$ being the
CS coupling constant which can take on any (real) value. 
The potential consists of two terms. The first term is the $D$-term of
$G'$ vector multiplet while the second term is that of the $U(1)$
vector multiplet. 

Although we have turned off the terms including fermions, one can
still smell the supersymmetric nature in the special relation between
the gauge and scalar coupling constants. 
The so-called Fayet-Iliopoulos (FI) parameter $\xi$ has been chosen 
to be positive which ensures stable supersymmetric (SUSY) vacua. 
In order to simplify the notation we define also the parameter
\beq v \equiv \sqrt{\frac{\xi}{N}} > 0 \ . \eeq
Note that the model has $SU(\NF)$ flavor symmetry.


If we completely discard the Higgs fields $H$ 
(i.e.~setting $H\equiv 0$) of the full Lagrangian (\ref{eq:LYMCSH}), 
the FI term in Eq.~(\ref{eq:LYMCSH}) can be absorbed by a constant
shift in $\phi^0$. Hence, the vacuum $\phi = 0$ is in the symmetric
phase where no symmetries are broken. Although the Higgs mechanism is
not at work, the vector multiplet acquires a topological mass which is
given by 
\beq
m_\kappa = \frac{\kappa e^2}{4\pi} \ , \quad
m_\mu = \frac{\mu g^2}{4\pi} \ .
\label{eq:topological_mass}
\eeq
Here $m_\kappa$ is the mass of the Abelian gauge fields $A_\mu^0$ as
well as the real scalar field $\phi^0$ while $m_\mu$ is the mass of
$G'$ gauge fields $A_\mu^a$ as well as the real adjoint fields
$\phi^a$.


In the following we will study the above described model in two
limits, namely $\kappa,\mu\to 0$ which reduces the model to the
Yang-Mills-Higgs (YMH) theory without the CS interactions and
$e,g\to\infty$ which reduces it to the ``pure'' Chern-Simons-Higgs (CSH)
theory without dynamical gauge fields. 
We will leave the intermediate case of both kinetic terms at finite
coupling, viz.~YMCS-Higgs theory for a companion
paper \cite{EtoGudnason}.

\subsection{Yang-Mills-Higgs theory \label{sec:model_ym}}

Taking the limit $\kappa,\mu\to 0$ eliminates the CS kinetic
term for the gauge fields and decouples the adjoint scalar field from
the $D$-terms. Hence the bosonic
Lagrangian density can now be written as\footnote{
This model can be trivially embedded in $3+1$ dimensions where the
vortex solutions describe strings instead of particles in which case
the supersymmetry allowed by the given potential is $\mathcal{N}=1$ in
$3+1$ dimensions. 
Either point of view is consistent with our discussion in the
following, i.e.~formally there will be no difference.} 
\begin{align}
\mathcal{L}_{\rm YMH} = &
-\frac{1}{4g^2}\left(F_{\mu\nu}^a\right)^2
+\frac{1}{2g^2}\left(\D_\mu\phi^a\right)^2
-\frac{1}{4e^2}\left(F_{\mu\nu}^0\right)^2
+\frac{1}{2e^2}\left(\p_\mu\phi^0\right)^2 \non
&+\Tr\left[\D_\mu H\left(\D^\mu H\right)^\dag\right] 
-\frac{g^2}{2}\left\{\Tr\left[HH^\dag t^a\right]\right\}^2
-\frac{e^2}{2}\left\{\Tr\left[\left(HH^\dag  - v^2 {\bf 1}_N\right)t^0\right]
\right\}^2
\ . \label{eq:LYMH}
\end{align}

In the case of $G'=SU(N)$ there is a unique Higgs vacuum where $G$ is
completely broken 
\beq H=v\mathbf{1}_N \ , \quad \phi=0 \ . \label{eq:Higgs_c+f_vacuum}
\eeq
The Higgs phase is in the color-flavor locking phase where the
diagonal global symmetry is preserved
\beq 
U(N) \times SU(\NF) \to SU(N)_{\rm c+f} \ . 
\eeq
The vacuum is gapped and there are no massless modes, while the  mass
spectrum is given by 
\beq
m_e = v e \ , \quad
m_g = v g \ .
\label{eq:higgs_mass}
\eeq
Thanks to supersymmetry, the vector multiplets and the chiral
multiplets have the same masses, viz.~the Abelian gauge fields
$A_\mu^0$, the real scalar field $\phi^0$ and the real part of the 
trace part of $H$ all have the same mass $m_e$. 
On the other hand, the $SU(N)$ gauge fields $A_\mu^a$, the real
adjoint fields $\phi^a$ and the real part of the traceless part of $H$
all have the mass $m_g$.

For $G'=SO(N),USp(N)$ the vacuum structure is much more complicated,
see Ref.~\cite{Eto:2008qw}. We will however, only consider the same
vacuum (\ref{eq:Higgs_c+f_vacuum}) for the reason that it preserves
the maximal global color-flavor locking symmetry
\beq G \times SU(\NF) \to G'_{\rm c+f}\ . \eeq
The masses of the vector multiplets remain the same as in the
$G'=SU(N)$ case. Those for the trace part and some of the traceless
part of $H$ are unchanged but the rest of the fields contained in $H$
become massless (i.e.~due to flat directions) since the number of
gauge fields is not sufficient for the Higgs mechanism to 
eat all the massless fields.

Let us consider topological solitons in this model.
We are interested in the vortex which is usually called the
non-Abelian vortex. It is a natural extension of the Nielsen-Olesen
vortex of the Abelian-Higgs model. 
Performing a Bogomol'nyi trick on the Hamiltonian under the assumption
that the configurations are static, i.e.~the energy in the
$\mathbb{C}$-plane 
\begin{align}
T = &\ \int_{\mathbb{C}} \bigg(
\frac{1}{2g^2}\left({F}_{12}^a - 
  g^2\Tr\left[H H^\dag t^a\right]\right)^2
+\frac{1}{2e^2}\left(F_{12}^0 - 
  e^2\Tr\left[\left(H H^\dag  -
  v^2\mathbf{1}_N \right) t^0\right]\right)^2 \label{eq:YMtension} \\
&\ \phantom{\int_{\mathbb{C}}\bigg[} 
+ \frac{1}{2g^2}(\D_i\phi^a)^2 + \frac{1}{2e^2}(\p_i\phi^0)^2
+\Tr\left[4 \left|\bar{\D} H\right|^2 - v^2  F_{12}
-i\epsilon^{ij}\p_i\left\{\left(\D_j H\right)H^\dag\right\}\right] \bigg)
\ , \nonumber
\end{align}
we can read off the BPS tension which is the lower bound
\beq T_{\rm BPS} = - v^2\int_{\mathbb{C}} \Tr\left[F_{12}\right]
= 2\pi v^2 N \nu = 2\pi\xi\nu > 0 \ .
\eeq 
Here $\nu$ is the Abelian winding number determined as
\beq 
\nu = - \frac{1}{2\pi N} \int_{\mathbb{C}} \Tr\left[F_{12}\right] 
  = \frac{k}{n_0} \ , \qquad
k\in\mathbb{Z}_{>0} \ , 
\label{eq:wind_ym}
\eeq
where $n_0$ is the center of the gauge group $\mathbb{Z}_{n_0}$ (see
Table \ref{tab:n0})\footnote{
$n_0$ is the greatest common divisor
(gcd) of the Abelian charges of all the $G'$ invariants which have a
non-zero VEV at infinity in the Higgs phase, see
Ref.~\cite{Eto:2008yi}.}.   
Finally, from Eq.~(\ref{eq:YMtension}) we have the BPS equations 
$(\phi^\alpha = 0)$ 
\beq
\bar{\D} H = 0, \qquad
F_{12}^\alpha = g_\alpha^2
\Tr\left[\left(H H^\dag - v^2 {\bf 1}_N\right) t^\alpha\right]\ ,
\label{eq:YMBPS1}
\eeq
where $g_\alpha$ stands for the gauge coupling constants ($g_0=e$ and
$g_a = g$) and $\alpha$ is not summed over. 
The non-Abelian part of the second equation is expressed in matrix
notation as 
\beq
\hat{F}_{12} = g^2  \left<HH^\dagger\right>_{G'} \ ,
\eeq
where we have introduced a bracket for a projection operation
\beq
\left<X\right>_{G'} \equiv \Tr\left[X t^a\right] t^a\ ,
\eeq
for an arbitrary $N\times N$ matrix $X$.

In the special case of $e=g$ we can simplify the second BPS equations
to the following combined equation 
\begin{align}
F_{12} = e^2 \left<HH^\dagger-v^2\mathbf{1}_N\right>_G \ ,
\label{eq:YMequalcouplingBPS}
\end{align}
where we have defined
\beq
\left<X\right>_G \equiv \Tr\left[X t^\alpha\right]t^\alpha 
= \frac{1}{2N}\Tr[X]{\bf 1}_N + \left<X\right>_{G'} \ .
\eeq
Note that $\left<X\right>_G = X/2$ holds if $X\in G$.

The above special case of $e=g$ is not only simple but also has
the advantage that solutions to the Abelian BPS equation automatically
solve the non-Abelian BPS equations. 
Hence, without solving the non-Abelian equations, we already
have a known class of solutions which are essentially Abelian
embeddings. 
All the minimally winding solutions are indeed in this
class. Of course, we have to solve the non-Abelian system for the
higher winding vortices which are out of the class. 

In particular, we can write the explicit form of $\left<X\right>_{G'}$
with $G'=SU(N),SO(N)$ and $USp(N)$ as 
\beq
\left<X \right>_{G'} = 
\left\{
\begin{array}{lcl}
\frac{1}{2}\left(X - \frac{1}{N}\Tr[X]{\bf 1}_N\right) \ , & & G'=SU(N) \ ,
\\
\frac{1}{4}\left(X - J^\dagger X^{\rm T} J\right) \ , & & G' = SO(N),USp(N) \ .
\end{array}
\right.
\eeq
For $U\in G'=SO(N),USp(N)$ we will use the invariant rank-two tensor
$J$ defined as
\beq
U^{\rm T} J U = J \ , \quad
J^\dag = \epsilon J \ ,\quad
J^\dag J = {\bf 1}_N \ ,
\label{eq:inv_tensor}
\eeq
with $\epsilon = +1$ for $SO(N)$ and $\epsilon = -1$ for $USp(N)$.
Throughout the paper we will adapt the basis in which $J$ is
\beq
J = 
\begin{pmatrix} 
0 & \mathbf{1}_M \\ 
\epsilon\mathbf{1}_M & 0
\end{pmatrix} \ , \label{eq:Jtensor_even}
\eeq
for $SO(2M)$ ($\epsilon=+1$) and $USp(2M)$ ($\epsilon=-1$) while for
$SO(2M+1)$ instead
\beq J = 
\begin{pmatrix} 
0 & \mathbf{1}_M & 0\\
\mathbf{1}_M & 0 & 0\\
0 & 0 & 1
\end{pmatrix} \ . \label{eq:Jtensor_odd}
\eeq

Let us give some examples of minimal winding solutions in the case of 
$G'=SU(4),SO(4)$ and $USp(4)$
\beq
SU(4)&:&\ 
\left\{
\begin{array}{l}
H = 
{\rm diag}\left(
H^{\rm ANO}_{k=1},v,v,v
\right)\\
F_{12} =
{\rm diag}\left(
F_{12,k=1}^{\rm ANO},0,0,0\right)
\end{array}
\right.,\\
SO(4),USp(4)&:&\ 
\left\{
\begin{array}{l}
H = 
{\rm diag}\left(
H^{\rm ANO}_{k=1},H^{\rm ANO}_{k=1},v,v
\right)
\\
F_{12} =
{\rm diag}
\left(
F_{12,k=1}^{\rm ANO},F_{12,k=1}^{\rm ANO},0,0\right)
\end{array}
\right. ,
\eeq
where $\left(H^{\rm ANO}_k,F^{\rm ANO}_{12,k}\right)$ stands for 
$k \in \mathbb{Z}_{\ge 0}$ winding coaxial-vortex solution in the
Abelian theory. 
Generic non-Abelian solutions in the $SU(4)$ and $USp(4)$ models 
can be generated by acting with the global symmetry of the system on
the above solutions. There exist further solutions which belong to the
different classes in the $SO(4)$ cases. 

The most significant feature of the non-Abelian vortex is the presence
of the so-called orientational zero-modes associated with the
spontaneous breaking of the color-flavor locked symmetry. 
In the above example of $G'=SU(4)$, one can easily see that
$SU(4)_{\rm c+f}$ obeyed in the vacuum ($r\to\infty$), is in fact
broken to $U(3)_{\rm c+f}$ at the center of the vortex where
$H_{k=1}^{\rm ANO} \to 0$. 
Thus the Nambu-Goldstone zero-modes of the
single non-Abelian vortex in the $G'=SU(4)$ theory are
\beq
{\cal M}_{SU(4)}^{k=1} = \mathbb{C} \times \mathbb{C}P^3 \ .
\eeq
Here the first factor, $\mathbb{C}$, corresponds to the position
zero-modes. 
Similarly, for the case of $G' = USp(4)$, the vortex breaks
$USp(4)_{\rm c+f}$ down to $U(2)_{\rm c+f}$. Hence, in this case the
zero-modes are 
\beq
{\cal M}_{USp(4)}^{k=1} = \mathbb{C} \times \frac{USp(4)}{U(2)} \ ,
\eeq
while for $G'=SO(4)$ there exists a topological charge in addition to
the vortex number, which is called the $\mathbb{Z}_2$ charge
\cite{Eto:2009bg}. The moduli space is then split into two copies of
the space \cite{Ferretti:2007rp,Eto:2009bg} given by the breaking of
$SO(4)_{\rm c+f}$ down to $U(2)_{\rm c+f}$ as 
\beq 
\mathcal{M}_{SO(4)}^{k=1} = \mathbb{C} \times \left[
  \left.\frac{SO(4)}{U(2)}\right|_{\mathbb{Z}_2=+} \cup 
  \left.\frac{SO(4)}{U(2)}\right|_{\mathbb{Z}_2=-}\right] \ . 
\eeq
In the case of $G'=SO(N)$, with $N>4$ the zero-modes are somewhat more 
elaborated, so we will not explain the details here, which however can
be found in Ref.~\cite{Eto:2009bg}.

\subsection{Chern-Simons-Higgs theory \label{sec:model_cs}}

The second limit under consideration is $e,g\to\infty$, i.e.~the
strong gauge coupling limit where the kinetic terms of the vector multiplets vanish.
Then the original Lagrangian density (\ref{eq:LYMCSH}) reduces to the 
CSH model (see Ref.~\cite{Gudnason:2009ut})
\begin{align}
\mathcal{L}_{\rm CSH} &=
-\frac{\mu}{8\pi}\epsilon^{\mu\nu\rho}
  \left(A_{\mu}^a\partial_\nu A_\rho^a  
  -\frac{1}{3}f^{abc}A_{\mu}^a A_{\nu}^b A_{\rho}^c\right)
-\frac{\kappa}{8\pi}
  \epsilon^{\mu\nu\rho} A_{\mu}^0\partial_\nu A_\rho^0 \non
&\phantom{=\ }
+\Tr\left[\D_\mu H\left(\D^\mu H\right)^\dag\right] 
- V_{\rm CSH}
 \ , \label{eq:LCSH}\\
V_{\rm CSH} &=  \Tr\left[\phi^2 HH^\dagger \right] \ .
\end{align}
The sixth-order scalar potential $V_{\rm CSH}$ is given by eliminating the 
adjoint scalar fields $\phi^\alpha$ from Eq.~(\ref{eq:LYMCSH}) by
\beq 
\phi^\alpha = \frac{4\pi}{\kappa_\alpha}
\Tr\left[\left(HH^\dag  - v^2 {\bf 1}_N \right)t^\alpha \right] \ ,\quad
\alpha=0,1,2,\ldots,\dim(G')\ ,
\label{eq:phi}
\eeq 
with $\kappa_0=\kappa$ and $\kappa_a=\mu$, for $a=1,2,\ldots,\dim(G')$, and
$\alpha$ is not summed over on the right-hand side.

There exist three types of vacua in this model:
\begin{itemize}
\item Symmetric phase:
One vacuum is in the completely symmetric phase (the CS
phase) where $\langle H\rangle=0$ and no symmetries are broken. The
vector multiplets are decoupled, so the Higgs fields are the only
dynamical degrees of freedom with the mass 
\beq
m_H = |\phi^0| = \frac{4\pi v^2}{\kappa} \sqrt{\frac{N}{2}} \ .
\eeq
\item Asymmetric phase:
There is also a vacuum in the Higgs phase $\langle H\rangle =
v\mathbf{1}_N$, where the gauge symmetry is completely broken. 
The mass of Higgs fields is the same as that of gauge fields due to
supersymmetry. 
The Abelian gauge fields $A_\mu^0$ and the real part of the trace part
of $H$ have the same mass $m_{\kappa\infty}$, while the $G'$ gauge
fields $A_\mu^a$, the real part of traceless part of $H$ have the mass
$m_{\mu\infty}$.
\beq
m_{\kappa\infty} = \frac{m_e^2}{m_\kappa} = \frac{4\pi v^2}{\kappa}
\ ,\quad 
m_{\mu\infty} = \frac{m_g^2}{m_\mu} = \frac{4\pi v^2}{\mu} \ . 
\label{eq:topological_mass2}
\eeq
The Higgs vacuum $\langle H\rangle=v{\bf 1}_N$ is unique if $G'=SU(N)$
but there exist flat directions in the cases of $G'=SO(N),USp(N)$ as we
have seen in the previous section.
\item Intermediate phases: 
In between there is a variety of partially broken phases. 
\end{itemize}

In the model with $G'=SU(N)$ and $\NF=N$, the vacua are labeled by an
integer $m=0,1,2,\ldots,N$ as follows
\beq
H^{(m)} = 
  \diag(\underbrace{v,v,\cdots,v}_m,\underbrace{0,\cdots,0}_{N-m})
  \ . \label{eq:mthpartiallybrokenphase}
\eeq
The vacuum $m=N$ is the full Higgs vacuum where the symmetry
$U(N)_{\rm c} \times SU(N)_{\rm f}$ is broken to $SU(N)_{\rm c+f}$. 
The vacuum $m=0$ is the unbroken phase where no symmetries are
broken, while in the case of $1\le m\le N-1$ we have the intermediate
vacua. In this case the symmetry is broken down to 
$U(N-m)_{\rm c} \times S[U(m)_{\rm c+f}\times U(N-m)_{\rm f}]$.
Note that the vacua with $m=0,N$ are unique but the rest are not.
The vacuum manifold is a complex Grassmannian manifold
given by $SU(N)_{\rm f}/S[U(m)_{\rm c+f}\times U(N-m)_{\rm f}]
\simeq Gr_{N,m}$.

The variety of the vacua results in a variety of
vortices which can be either topological or non-topological. 
In order to derive the BPS equations, we again perform a Bogomol'nyi
trick on the tension of the vortex 
\begin{align}
T =&\ \int_{\mathbb{C}}\Tr\left[
\left|\left(\D_0  
- i\phi  \right)H\right|^2
+ 4\left|\bar{\D}H\right|^2 
- v^2 F_{12} 
-i\epsilon^{ij}\p_i\left\{\left(\D_j
  H\right)H^\dag\right\} \right]\ , 
\end{align}
where $\phi$ is given by Eq.~(\ref{eq:phi}).
The tension for the BPS saturated vortices is given by
\beq T_{\rm BPS} = - v^2 \int_{\mathbb{C}} \Tr [F_{12}] 
= 2\pi Nv^2\nu > 0 \ . \eeq
where $\nu$
is the $U(1)$ winding number 
\beq
\nu = - \frac{1}{2\pi\sqrt{2N}}\int_{{\mathbb C}} F^0_{12} = 
\left\{
\begin{array}{ccl}
\frac{k}{n_0} & & \text{for topological solitons}\\
\frac{k+\alpha}{n_0} & & \text{for non-topological solitons}\\
\frac{k}{n_0} + \frac{k'+\alpha}{n_0} & & \text{for topological and non-topological solitons}
\end{array}
\right.,
\label{eq:wind_cs}
\eeq
where $k,k'$ are integers and $\alpha$ is a real number. In the case of
Abelian non-topological solitons, the integer part $k$ is limited as 
$\alpha>k+2$, see Ref.~\cite{Khare:1991zt}. 
For non-Abelian cases, no conditions for $\alpha$ and $k$ are in
general known yet. 
But only in the special case of $U(N)$ gauge group
where the coupling constants are equal $\kappa = \mu$
and for the diagonal solutions, we can easily extend the Abelian
results as 
$\alpha_i > k_i + 2$. Here $\alpha_i$ ($\alpha = \sum_i^N\alpha_i$) 
and $k_i$  ($k = \sum_i^Nk_i$) are associated with 
the non-topological and topological flux of the $i$-th $U(1)$ subgroup
of $U(1)^N \subset U(N)$. 

Let us furthermore define the magnetic flux as follows 
\beq
\Phi \equiv 2\pi N \nu
\ . \label{eq:Phi_def}
\eeq

The BPS equations in this case read
\begin{align}
\bar{\D} H = 0 \ ,\qquad 
\D_0 H = i \phi H \ ,\label{eq:CSBPS1}
\end{align}
which however due to the presence of electric charge density in the 
vortex have to be accompanied by the Gauss law ($\alpha$ not summed
over) 
\begin{align}
F_{12}^\alpha =
-\frac{i4\pi}{\kappa_\alpha}\Tr\left[\left((\D_0 H)H^\dag -  H\left(\D_0 H\right)^\dag \right)
 t^\alpha \right] \ . 
\label{eq:Gausslaw}
\end{align}
Combining the equations (\ref{eq:CSBPS1}) and (\ref{eq:Gausslaw}) we
obtain the following system
\beq
\bar{\D}H = 0 \ , \qquad
F_{12}^\alpha = \frac{4\pi}{\kappa_\alpha}
  \Tr\left[\left\{\phi, HH^\dagger\right\} t^\alpha\right] \ .
\label{eq:NACSmastersystem1}
\eeq
This system is valid for any simple group $G'$.
In matrix notation, we can explicitly write down the second equation
as 
\begin{align}
F_{12}^0t^0 &= 
  \frac{8\pi^2}{N^2\kappa^2}\Tr\left[H H^\dag - v^2 {\bf 1}_N\right]
  \Tr\left[HH^\dag\right]{\bf 1}_N
+\frac{16\pi^2}{N\kappa\mu} 
  \Tr\left[\left<H H^\dag\right>_{G'}H H^\dag\right]{\bf 1}_N \ ,
  \label{eq:NACSmaster1}\\
\hat{F}_{12} &= \frac{16\pi^2}{N\mu\kappa} 
  \Tr\left[HH^\dagger-v^2{\bf 1}_N\right]\left<HH^\dagger\right>_{G'}
+\frac{32\pi^2}{\mu^2}
  \left<\left<H H^\dag\right>_{G'}H H^\dag\right>_{G'} \ ,
\label{eq:NACSmaster2}
\end{align}
where we have used the following identity
\begin{align}
\left<X\left<X\right>\right>_{G'} &=
\Tr\left[X t^b t^a\right]\Tr\left[X t^b\right] t^a \non &= 
\Tr\left[X t^a t^b\right]\Tr\left[X t^b\right] t^a +
i f^{b a c}\Tr\left[X t^c\right]\Tr\left[X t^b\right] t^a \non &= 
\left<\left<X\right>X\right>_{G'} \ ,
\label{eq:bracketidentity}
\end{align}
due to the fact that $f^{b a c} = - f^{c a b}$ in the basis
(\ref{eq:liebasis}).
In the equal coupling case $\kappa=\mu$ we can simplify the BPS
equations (\ref{eq:NACSmaster1}) and (\ref{eq:NACSmaster2}) 
to the following combined equation
\beq
F_{12} = \frac{32\pi^2}{\kappa^2}
\left<\left<H H^\dag-v^2{\bf 1}_N\right>_G H H^\dag\right>_G \ .
\label{eq:CSequalcouplingBPS}
\eeq

Now we have seen the differences between the two different BPS systems
in Sec.~\ref{sec:model_ym} and Sec.~\ref{sec:model_cs}.
The equation for the Higgs fields $H$ (Eqs.~(\ref{eq:YMBPS1}) and
(\ref{eq:NACSmastersystem1})), is common. 
The difference resides only in the flux equations
(\ref{eq:YMequalcouplingBPS}) and (\ref{eq:CSequalcouplingBPS}), which
will eventually become the master equations after applying the moduli 
matrix method in Sec.~\ref{sec:modulimatrix}.
We want to study the zero-modes of the BPS solutions to these
equations in the subsequent sections.

\subsubsection{The Abelian Chern-Simons solitons}

Let us recall the solitons in the minimal model by choosing $N=1$ and
$\NF=1$, namely the Abelian CSH model
\cite{Jackiw:1990pr}. The BPS equations are 
\beq
\bar{\D} H = 0 \ ,\qquad
\frac{1}{\sqrt{2}}F_{12}^0 = 
  \frac{8\pi^2}{\kappa^2}\left(|H|^2 - v^2\right)|H|^2 \ .
\eeq
It is known that there are two kinds of solitons: i) the topological
solitons and ii) the non-topological solitons. 
Let $(H^{\rm TP},F_{12}^{\rm TP})$ be a topological solution
and $(H^{\rm NTP},F_{12}^{\rm NTP})$ be a non-topological solution
(several numerical solutions have been obtained in
Ref.~\cite{Jackiw:1990pr}).  
The topological solitons live in the Higgs vacuum while the
non-topological solitons live in the unbroken (CS)
vacuum. Therefore, the asymptotic behavior of the Higgs field is  
\beq
|H^{\rm TP}| \to v \ ,\quad 
|H^{\rm NTP}| \to 0 \ ,\qquad \text{as}\quad r \to \infty \ .
\eeq
For topological reasons $H^{\rm TP}$ must vanish at the center of the
vortex but this is not the case for the non-topological soliton
\beq
H^{\rm TP} \to 0 \ ,\quad 
H^{\rm NTP} \to v' \ ,\qquad \text{as}\quad r \to 0 \ ,
\eeq
where $v'\in (0,v)$ is a constant.
The topological soliton has a quantized magnetic flux
(\ref{eq:Phi_def}) whereas the flux of the non-topological one is a
continuum 
\beq
\Phi^{\rm TP} =  2\pi k \ ,\quad 
\Phi^{\rm NTP}=  2\pi (k + \alpha) \ ,
\eeq
with $k \in\mathbb{Z}_{\geq 0}$ and $\alpha > k + 2$. 
For both topological and non-topological vortices, the integer $k$
corresponds to the number of the zeros of the Higgs field $H$.

It turns out that the dimension of the moduli space of the
Abelian CS solitons is
\beq
\dim_{\mathbb{C}} {\cal M}_{\rm ACS} = k + \hat\alpha -1 \ .
\label{eq:dim_ACS}
\eeq
Here $\hat{\alpha}$ the integer-part of the real number $\alpha$
\cite{Jackiw:1990pr}.

\subsubsection{The $U(2)$ Chern-Simons solitons \label{sec:u2_cs}} 

To describe some of the characteristic features of the non-Abelian
extension of the CS solitons, we will discuss the simple
example of $G'=SU(2)$ with $\NF=2$ in this subsection. 
We also set $\kappa=\mu$ so that the Abelian solutions automatically 
solve the non-Abelian equations
\beq
\bar{\D} H = 0 \ , \qquad
F_{12} = \frac{8\pi^2}{\kappa^2}
  \left(HH^\dag - v^2{\bf 1}_2\right)HH^\dag \ .
\eeq
Let us first figure out the vacuum structure of the model.
There exist three different vacua: (2) the Higgs vacuum, (0) the
unbroken (CS) vacuum and (1) intermediate vacua 
\beq
H^{(2)}= 
\begin{pmatrix}
v & 0 \\
0 & v
\end{pmatrix} \ , \quad
H^{(1)} = 
\begin{pmatrix}
v & 0 \\
0 & 0
\end{pmatrix} \ , \quad
H^{(0)} = 
\begin{pmatrix}
0 & 0 \\
0 & 0
\end{pmatrix} \ .
\label{eq:vac_u2_cs}
\eeq
The symmetry of the vacua is $SU(2)_{\rm c+f}$ in the Higgs vacuum,
$U(1)_{\rm c} \times U(1)_{\rm c+f}$ in the intermediate vacuum and
finally $U(2)_{\rm c} \times SU(2)_{\rm f}$ in the unbroken vacuum. 
As already mentioned, the Higgs and unbroken vacua are unique but the
intermediate vacuum possesses a continuum. By acting with 
the $SU(2)_{\rm f}$ flavor symmetry, one can transform the above given
configuration into a generic point 
\beq
H^{(1)} = v 
\begin{pmatrix}
\phi_1 & \phi_2\\
0 & 0
\end{pmatrix} \ , \qquad
|\phi_1|^2 + |\phi_2|^2 = 1 \ .
\eeq
Since the overall phase of $\phi_1,\phi_2$ is gauged, the vacuum
manifold is $SU(2)/U(1) \simeq \mathbb{C}P^1$. One can exchange the
first and second rows in the matrix, which however is $SU(2)_{\rm c}$
gauge equivalent to the given form. 

In the Higgs vacuum we can construct the topological vortex as follows 
\beq
H = 
\begin{pmatrix}
H^{\rm TP} & 0 \\
0 & v
\end{pmatrix}
\to
\left\{
\begin{array}{ccl}
{\rm diag}\left(0,v\right) \ , & & \text{as}\quad r\to 0 \ ,\\
{\rm diag}\left(v,v\right) \ , & & \text{as}\quad r\to\infty \ .
\end{array}
\right.
\eeq
Clearly, the color-flavor symmetry $SU(2)_{\rm c+f}$ respected by the
vacuum is spontaneously broken to $U(1)_{\rm c+f}$ at the center of
the vortex. This yields Nambu-Goldstone zero-modes which are the 
so-called orientational zero-modes in addition to the position
(translational) zero-modes. 
Thus the minimal topological vortex carries the zero-modes \cite{Aldrovandi:2007nb}
\beq
{\cal M}_{k=1}^{\rm TP} = \mathbb{C} \times \mathbb{C}P^1 \ .
\eeq

In the unbroken phase, we can construct the non-topological solitons
as follows 
\beq
H = 
\begin{pmatrix}
H^{\rm NTP} & 0 \\
0 & 0
\end{pmatrix}
\to
\left\{
\begin{array}{ccl}
{\rm diag}\left(v',0\right) \ , & & \text{as}\quad r\to 0 \ , \\
{\rm diag}\left(0,0\right) \ , & & \text{as}\quad r\to\infty \ .
\end{array}
\right.
\eeq
The symmetry of the vacuum $U(2)_{\rm c} \times SU(2)_{\rm f}$ is
spontaneously broken to $U(1)_{\rm c} \times U(1)_{\rm c+f}$ at the
center of the soliton. The spontaneous breaking of the global symmetry 
leads to the Nambu-Goldstone modes $SU(2)_{\rm f}/U(1)_{\rm c+f} \simeq \mathbb{C}P^1$. 
For simplicity, let us choose the minimal
non-topological soliton (corresponding to $\hat{\alpha}=2$)
which in the Abelian case has only translational zero-modes. 
Clearly, with the above symmetry argument it indeed carries the same 
zero-modes as the topological soliton 
\beq
{\cal M}_{\hat \alpha =2}^{\rm NTP} = \mathbb{C} \times \mathbb{C}P^1
\ ,
\eeq
where the first $\mathbb{C}$ corresponds to the translation
zero-mode. 
We will denote this the non-Abelian non-topological soliton. 

The intermediate vacua are interesting. We can put either topological
or non-topological solitons. Let us first construct the topological
soliton. A naive way is simply to embed the Abelian topological
solution as
\beq
H = 
\begin{pmatrix}
H^{\rm TP} & 0 \\
0 & 0
\end{pmatrix}
\to
\left\{
\begin{array}{ccl}
{\rm diag}\left(0,0\right) \ , & & \text{as}\quad r\to 0 \ , \\
{\rm diag}\left(v,0\right) \ , & & \text{as}\quad r\to\infty \ . 
\end{array}
\right.
\label{eq:naive_stp}
\eeq
Since the symmetry at the vortex center is larger than that of the
vacuum, there are no orientational zero-modes. However, this is not
the most general solution. The generic solutions can be obtained by
embedding not the Abelian solution with $\NF=1$ but the Abelian
solutions with $\NF=2$, hence the semi-local Abelian vortex solution
which have been studied in Ref.~\cite{Khare:1992qr}. 
Let $(H^{\rm STP}_1,H^{\rm STP}_2)$ be the fields of the semi-local
vortex solution in the Abelian theory. Then the generic solutions of
the $U(2)$ model can be written as follows
\beq
H = 
\begin{pmatrix}
H^{\rm STP}_1 & H^{\rm STP}_2 \\
0 & 0
\end{pmatrix}
\to
\left\{
\begin{array}{ccl}
\begin{pmatrix}
0 & v'\\
0 & 0
\end{pmatrix} \ , 
& & \text{as}\quad r\to0\\
\begin{pmatrix}
v & 0\\
0 & 0
\end{pmatrix} \ , 
& & \text{as}\quad r\to\infty
\end{array}
\right.,
\eeq
with $v'$ being a constant which should be determined by the equations
of motion. 
It is known that the semi-local vortex $(H^{\rm STP}_1,H^{\rm STP}_2)$ 
has a complex free parameter, say $a \in \mathbb{C}$. We can identify
$|a|$ with the transverse size of the vortex while the phase is the
relative $U(1)$ phase between the first and second Higgs elements. 
When we send $a$ to zero, the generic solution becomes the naive
embedding solution (\ref{eq:naive_stp}). Thus we conclude that the
moduli space of the topological solitons in the intermediate vacua is 
\beq
{\cal M}_{k=1}^{\rm STP} = \mathbb{C} \times \mathbb{C} \ .
\eeq
The first $\mathbb{C}$ is the position and the second corresponds to
the semi-local size modulus $a$.
This embedding solution is easily extended to the $U(N)$ case.
As explained before, there are $N-1$ different kinds of intermediate vacua,
see Eq.~(\ref{eq:mthpartiallybrokenphase}).
Obviously, in the vacua with $m=1$ and $N-m=1$ we can embed the
Abelian semi-local vortex of the topological and non-topological kind,
respectively. 
In the rest vacua with $m>1$ and $N-m>1$ we can have non-Abelian
semi-local vortices. 
So far, the semi-local vortices have been observed only in theories 
in which the number of flavors is greater than that of colors 
($\NF > N$), in the case of $G'=SU(N)$, see
Refs.~\cite{Lozano:2007yz,Buck:2009pd}. 
Here we first show that the semi-local vortices also exist in the
model with $\NF = N$. Furtheremore, if one likes, one can also
consider systems with $\NF < N$, for instance $\NF=2$ and $N=3$, and
construct the semi-local vortex there by choosing an appropriate
vacuum. \footnote{In $G'=SO,USp$ theories it is however different due
  to the already mentioned complicated vacuum structure, and the
  existence of the flat directions allows for semi-local vortices even 
  for the minimal number of flavors in order to put the theory on the
  Higgs branch \cite{Eto:2008qw,Eto:2008yi,Gudnason:2010yy}.}

There exists also a semi-local non-topological soliton in the Abelian
theory with $\NF=2$, see Ref.~\cite{Khare:1992qr}. 
Let $(H^{\rm SNTP}_1,H^{\rm SNTP}_2)$ be a minimal solution (which
corresponds to $\hat{\alpha}=2$). Then we can embed this into the
non-Abelian model as follows  
\beq
H = 
\begin{pmatrix}
v & 0 \\
H^{\rm SNTP}_1 & H^{\rm SNTP}_2
\end{pmatrix}
\to
\left\{
\begin{array}{ccl}
\begin{pmatrix}
v & 0\\
v' & v''
\end{pmatrix}
& & \text{as}\quad r\to 0 \ , \\
\begin{pmatrix}
v & 0\\
0 & 0
\end{pmatrix}
& & \text{as}\quad r\to\infty \ .
\end{array}
\right.
\eeq
The solution has also one complex free parameter so the moduli space
is again 
\beq
{\cal M}_{\hat \alpha =2}^{\rm SNTP} = \mathbb{C} \times \mathbb{C} \ .
\eeq

In Sec.~\ref{sec:modulimatrix}, we will study generic solutions and 
explicitly see the structure of moduli space by using the moduli matrix method.

\section{The index theorem\label{sec:indextheorem}}

\subsection{Calculation of the index\label{sec:indexcalc}}

Let us first make a general calculation of the index.
Later we will apply the results to the YMH and
the CSH models considered above. 

Suppose the zero-modes $\eta$ of a set of BPS equations are described
by a Dirac-type equation 
\beq 
\mathbb{D}\eta = 0 \ ,\qquad
\eta = 
\begin{pmatrix}
\eta_{\rm f}\\
\eta_{\rm a}
\end{pmatrix}
\ .
\label{eq:dirac}
\eeq
where $\eta$ takes a value in a vector-space generally described by
matrices with $\eta_{\rm f}$ being an $\NF \times N$ matrix while 
$\eta_{\rm a}$ is an $N\times N$ matrix.
The inner product on this vector space is defined as
\beq 
\left(\iota, \eta\right) 
\equiv \int_{\rm C} \Tr\left[\iota_{\rm f}^\dagger \eta_{\rm f}
+ \iota_{\rm a}^\dag \eta_{\rm a}\right] \ , \label{eq:innerprod} 
\eeq
while the adjoint operator of $\mathbb{D}$ is denoted by
$\mathbb{D}^\dag$ and is defined as usual
$\left( \iota, \mathbb{D}\eta \right) = 
\left( \mathbb{D}^\dag \iota, \eta \right)$.

In this section, we will calculate the index of the operator
$\mathbb{D}$ of the Callias type \cite{Callias:1977kg} (i.e.~the
generalization of the Atiyah-Singer theorem to non-compact manifolds), 
which is formally 
\begin{align}
\mathcal{I} = \dim \left(\ker\,\mathbb{D}\right) -
\dim \left(\ker\,\mathbb{D}^\dag\right) 
= \dim(\ker\,\mathbb{D}^\dag\mathbb{D}) - \dim(\ker\,\mathbb{D}\mathbb{D}^\dag)\ ,
\end{align}
where we have used that the kernel of $\mathbb{D}$
($\mathbb{D}^\dagger$) is the same as $\mathbb{D}^\dag\mathbb{D}$
($\mathbb{D}\mathbb{D}^\dag$). We will however calculate a slightly
different index defined by\footnote{We will adopt the notation where
  $\tr$ denotes trace over states as well as over the matrices while
  $\Tr$ denotes only a matrix trace.}
\begin{align} 
\hat{\mathcal{I}}(M^2) &= 
\tr\left(\frac{M^2}{\mathbb{D}^\dag\mathbb{D}+M^2}\right)
-\tr\left(\frac{M^2}{\mathbb{D}\mathbb{D}^\dag+M^2}\right) \ ,
\end{align}
which clearly gives back the index $\mathcal{I}$ in the limit 
$M^2\to 0$ if the system possesses a mass gap, but in most cases it
will turn out to be independent\footnote{This can be proven for a
  compact manifold and the result proves to be the same if the fields
  are sufficiently well-behaved at spatial infinity.} 
of $M^2$, hence it will prove useful to evaluate the index in the
limit $M^2\to\infty$. Subtleties arise in the case where the continuum 
part of the spectrum extends down to zero and one has to be more
careful \cite{Kiskis:1977vh}. This is because an unbounded resonance 
gives a certain contribution to the index. In such cases, we have to
subtract the continuum part from $\hat{\cal I}$ \cite{Jackiw:1990pr}.

Let us now make some general statements for the form of the operators
$\mathbb{D},\mathbb{D}^\dag$. First of all, we decompose the Dirac
operators as 
\begin{align}
\mathbb{D} = i D + K \ , \quad
\mathbb{D}^\dag = i D^\dag + K^\dag \ , 
\end{align}
where $D$ includes only the differential operators $\p,\bar\p$ and
$K,K^\dagger$ contain the remaining parts. This yields 
\beq
\mathbb{D}^\dag\mathbb{D} = \Delta - L_1,\qquad
\mathbb{D}\mathbb{D}^\dag = \Delta - L_2,
\eeq
where we have defined 
\beq
\Delta &\equiv& - D^\dagger D = - D D^\dagger\ ,\\
L_1 &\equiv& - i D^\dag K - i K^\dag D - K^\dag K \ ,\\
L_2 &\equiv& - i D K^\dag - i K D^\dag - K K^\dag \ .
\eeq
We assume that $\Delta$ is diagonal and composed by Laplacians, while 
$L_{1,2}$ contain only linear differential operators. Expanding in
terms of the large mass-squared parameter, we obtain 
\begin{align}
\frac{M^2}{\mathbb{D}^\dag\mathbb{D}+M^2} = M^2
\left[P^{-1}
+P^{-1}L_1P^{-1}
+P^{-1}L_1P^{-1}L_1P^{-1}
+\ldots\right] \ , \non
\frac{M^2}{\mathbb{D}\mathbb{D}^\dag+M^2} = M^2
\left[P^{-1}
+P^{-1}L_2P^{-1}
+P^{-1}L_2P^{-1}L_2P^{-1}
+\ldots\right] \ ,
\end{align}
where we have defined $P \equiv \Delta + M^2$.
All terms beyond the second order in $L_i$ will vanish in the large
mass limit $M^2\to\infty$. 

Now we will pursue a technique used in Abelian systems \cite{Lee:1991td} 
to simplify the calculation, which will demonstrate
that the terms $K^\dag K$ and $K K^\dag$ do not contribute to the
index. 
We can express the index in
the following four terms as 
\beq
\hat {\cal I} = {\cal I}_1 + {\cal I}_2 + {\cal I}_3 + {\cal I}_4 \ ,
\label{eq:fullindex}
\eeq
with
\beq
{\cal I}_1 &\equiv& \lim_{M^2\to\infty} i M^2 \tr
\left[
P^{-1} \left(
D K^\dag - K^\dag D - D^\dag K + K D^\dag
\right)
P^{-1}
\right]\ ,\\
{\cal I}_2 &\equiv& \lim_{M^2\to\infty} M^2 \tr
\left[
P^{-1} \left(
KK^\dagger - K^\dagger K
\right)
P^{-1}
\right]\ ,\\
{\cal I}_3 &\equiv& - \lim_{M^2\to\infty} M^2 \tr
\left[
P^{-1}(D^\dagger K + K^\dagger D) P^{-1} (D^\dagger K + K^\dagger D) P^{-1}
\right]\ ,\\
{\cal I}_4 &\equiv& \lim_{M^2\to\infty} M^2 \tr
\left[
P^{-1}(DK^\dag + KD^\dag) P^{-1} (D K^\dag + K D^\dag) P^{-1}
\right]\ .
\eeq
The trick now is that ${\cal I}_2 + {\cal I}_3 + {\cal I}_4 = 0$ 
which can be easily proven by making use of
\beq
\left[P^{-1},D\right] = \left[P^{-1},D^\dagger \right] = 0,\quad
DP^{-1}D^\dagger = D^\dagger P^{-1} D = -{\bf 1} + M^2 P^{-1}.
\eeq
Hence, we can write the index simply as
\begin{align}
\hat{\mathcal{I}} = \mathcal{I}_1.
\label{eq:indexsimplified}
\end{align}
This formula is considerably simpler than the starting point
(\ref{eq:fullindex}).

Let us calculate the index for a class of operators which includes 
our models 
\begin{align}
\mathbb{D} &= 
\begin{pmatrix}
i\bar{\D}_{\rm fund}^{\rm T} & K_1 \\
K_2 & i\D_{\rm adj}^{\rm T}
\end{pmatrix} \ , 
  \quad
\bar{\D}_{\rm fund}^{\rm T}  = \bar{\p} + i \circ \bar{A}^{\rm T} \ ,
  \quad
\D_{\rm adj}^{\rm T} = \p - i \left[A^{\rm T},\circ\right] \ .
\label{eq:operatorclass}
\end{align}
where the gauge fields $A,\bar{A}$ of the gauge group $G$ is a given
background configuration. 
The Dirac operator acts on fields of the following form
\beq
\eta = \left(
\begin{array}{c}
\delta\!H^{\rm T} \\
\delta\!\bar{A}^{\rm T}
\end{array}
\right) \ ,
\eeq
where $\delta\!H$ is an $N\times \NF$ matrix-valued field in the
fundamental representation of $G$ whereas $\delta\!\bar{A}$ is an 
$N\times N$ matrix-valued field in the adjoint representation of
$G$. $K_1$ (i.e.~an $\NF\times N$ matrix) and $K_2$ 
(i.e.~an $N\times \NF$ matrix) need not be fixed yet, so we leave them
as variables until we will need the specific forms.
It follows that with $\mathbb{D} = iD + K$ 
\begin{align}
D =
\begin{pmatrix}
\mathbf{1}_{\NF}\bar{\p} & 0 \\
0 & \mathbf{1}_N \p
\end{pmatrix} \ , \quad
K = 
\begin{pmatrix}
- \circ \bar{A}^{\rm T} & K_1 \\
K_2 & \left[A^{\rm T},\circ\right] 
\end{pmatrix} \ ,
\end{align}
and its adjoint operator $\mathbb{D}^\dag = i D^\dag + K^\dag$ is defined by
\begin{align}
D^\dag = 
\begin{pmatrix}
\mathbf{1}_{\NF}\p & 0 \\
0 & \mathbf{1}_N \bar{\p}
\end{pmatrix},\qquad
K^\dag = 
\begin{pmatrix}
- \circ A^{\rm T} & K_2^\dag \\
K_1^\dag  & \left[\bar{A}^{\rm T},\circ\right] 
\end{pmatrix} \ ,
\end{align}
where $K_{1,2}^\dag$ are the adjoint operators of $K_{1,2}$.
The Laplacian operator reads
\beq \Delta = -D^\dag D = - D D^\dag = - \bar{\p}\p \ .
\label{eq:laplacianoperator}
\eeq
Now by a simple calculation we can show that
\begin{align}
D K^\dag - K^\dag D - D^\dag K + K D^\dag =
\begin{pmatrix}
\circ \left(\p\bar{A}^{\rm T}-\bar{\p}A^{\rm T}\right) & * \\
* & \left[\p\bar{A}^{\rm T} - \bar{\p}A^{\rm T},\circ\right]
\end{pmatrix} \ ,
\label{eq:effectivematrix}
\end{align}
independently of $K_{1,2}$.\footnote{Even though the index theorem
  calculation does not depend on the constants $K_{1,2}$, they are
  needed to derive the vanishing theorem, see below.} 

We can now easily calculate the index by means of
Eq.~(\ref{eq:indexsimplified}) 
\begin{align}
\lim_{M^2\to\infty}\hat{\mathcal{I}} &= 
i M^2 \int d^2x \; \tr\left\langle x\left|
P^{-1} \left(
D K^\dag - K^\dag D - D^\dag K + K D^\dag
\right)
P^{-1}
\right| x\right\rangle \non
&=
-\sum^{\NF} \int_{\mathbb{C}} \; i \;\Tr
\left(\bar{\p}A - \p\bar{A}\right)
\int \frac{d^2k}{\left(2\pi\right)^2}
\frac{M^2}{\left(\frac{1}{4}k^2+M^2\right)^{2}} \non
&= \NF N \nu \ , 
\label{eq:index}
\end{align}
which turns out to be independent of $M^2$ and 
the Abelian winding number $\nu$ is defined in Eqs.~(\ref{eq:wind_ym}) and
(\ref{eq:wind_cs}), for YMH and CSH theories, respectively.

The result (\ref{eq:index}) is quite general and can be applied to 
all the solitons under consideration in this paper. 
Namely, the index (\ref{eq:index}) is applicable to all the models
considered in this paper: YMH and CSH
theories with a general gauge group 
\beq G = [U(1) \times G']/\mathbb{Z}_{n_0} \ . \nonumber  \eeq

Next we want to show that the kernel of $\mathbb{D}^\dag$ is indeed
vanishing. This statement is usually denoted the vanishing theorem,
and we will need to compute those case by case. The applicability of
the vanishing theorem makes the index equal to the number of
zero-modes of the system described by the operator $\mathbb{D}$.


\subsection{Yang-Mills-Higgs theory \label{sec:index_YM}}


\subsubsection{The index for Yang-Mills vortices}

The index theorem in the case of $G'=SU(N)$ YM theory was already made in 
Ref.~\cite{Hanany:2003hp} while it was made with
a general group in Ref.~\cite{Eto:2009bg}, hence we will just briefly
review the calculation here as we will need it later. 

In what follows, we set $e=g$, in order to make the expressions more 
compact. Our starting point is the BPS equations (\ref{eq:YMBPS1}) and
(\ref{eq:YMequalcouplingBPS}). 
Let $\{H,H^\dag,A,\bar{A}\}$ be a BPS solution.
Then we consider small fluctuations around it as
$H \to H + \delta\!H$, $H^\dag \to H^\dag + \delta\!H^\dag$, 
$A \to A + \delta\!A$ and $\bar A \to \bar A + \delta\!\bar{A}$. 
The fluctuations obey the following linearized BPS equations
\begin{align}
i\bar{\D}_{\rm f}\,\delta\!H - \delta\!\bar{A}\, H &= 0 \ , 
  \label{eq:YMfluc1} \\
i\bar{\D}_{\rm a}\,\delta\!A - i\D_{\rm a}\,\delta\!\bar{A} &=
\frac{e^2}{2}\left(\left<\delta\!H H^\dagger \right>_G 
+ \left<H\delta\!H^\dagger \right>_G \right)\ ,
\label{eq:YMfluc2}
\end{align}
where we have used that 
$\delta\!F_{12}=2 i\left(\bar{\D}_{\rm a}\,\delta\!A
-\D_{\rm a}\,\delta\!\bar{A}\right)$. 
We need to introduce a gauge fixing in order not to count non-physical
degrees of freedom and it will prove convenient to choose \cite{Hanany:2003hp}
\beq 
i\bar{\D}_{\rm a}\,\delta\!A + i\D_{\rm a}\,\delta\!\bar{A} =
\frac{e^2}{2} 
\left(-\left<\delta\!H H^\dagger \right>_G 
  +\left<H\delta\!H^\dagger \right>_G \right) \ . 
\label{eq:gauss}
\eeq
Combining Eqs.~(\ref{eq:YMfluc2}) and (\ref{eq:gauss}), we can now
write these equations in the form of Eq.~(\ref{eq:dirac}) with the
following Dirac operator 
\begin{align}
\mathbb{D} = 
\begin{pmatrix}
i\bar{\D}_{\rm f}^{\rm T} & - H^{\rm T} \\
\frac{e^2}{2}\left<\bar{H}\circ\right>_G & i\D_{\rm a}^{\rm T}
\end{pmatrix} \ , \quad
\mathbb{D}^\dag = 
\begin{pmatrix}
i\D_{\rm f}^{\rm T} &  \frac{e^2}{4}H^{\rm T} \\
-2\left<\bar{H}\circ\right>_G & i\bar{\D}_{\rm a}^{\rm T}
\end{pmatrix}\ ,\quad
\eta = 
\begin{pmatrix}
\delta\!H^{\rm T} \\
\delta\!\bar{A}^{\rm T}
\end{pmatrix} \ ,
\label{eq:diracYM}
\end{align}
where $\bar{H}$ is just the complex conjugate of $H$ while $H^\dag$ is
the Hermitian conjugate as usual. 
Since this operator is in the class of operators
(\ref{eq:operatorclass}), we can immediately apply the result
of the index (\ref{eq:index}).
From Eqs.~(\ref{eq:wind_ym}) and (\ref{eq:index}), we find the index
\beq
{\cal I} = \lim_{M^2\to\infty} \hat{{\cal I}} = \frac{\NF N k}{n_0} \ .
\label{eq:res_index_ym}
\eeq


\subsubsection{Vanishing theorem for Yang-Mills
  vortices \label{sec:vanish_YM}} 

In order to establish that the index does in fact correspond to the
number of bosonic zero-modes of the BPS vortex solutions, we need to
demonstrate that the dimension of the kernel of the adjoint operator
is indeed zero
\begin{align}
\dim \left(\ker\mathbb{D}^\dag\right) = 0 \ .
\label{eq:vanishing_th}
\end{align}
The question is if there exist normalizable zero-modes $\chi$ for
$\mathbb{D}^\dag$ 
\beq
\mathbb{D}^\dag \chi = 0 \ , \qquad
\chi = 
\begin{pmatrix}
\chi_{\rm f}^{\rm T}\\
\chi_{\rm a}^{\rm T}
\end{pmatrix} \ ,
\eeq
where $\chi_{\rm f}^{\rm T}$ is an $\NF$-by-$N$ matrix and 
$\chi_{\rm a}^{\rm T}$ is an $N$-by-$N$ matrix and they obey the
following equations 
\beq
\D_{\rm f}\chi_{\rm f} = \frac{i e^2}{4}\chi_{\rm a} H \ , \qquad
\bar\D_{\rm a}\chi_{\rm a} = -i2 \left<\chi_{\rm f}H^{\dag}\right>_G
\ , \label{eq:fluc_equations_YM}
\eeq
which can be combined into the following equation
\beq
\D_{\rm a}\bar\D_{\rm a} \chi_{\rm a} 
  = \frac{e^2}{2} \left< \chi_{\rm a}H H^\dag\right>_G \ .
\label{eq:schrodinger_YM}
\eeq

In the Abelian case this is simply the Schr\"odinger-type equation 
\beq
\left(-\p\bar\p + \frac{e^2}{4}|H|^2\right)\chi_{\rm a} = 0 \ .
\eeq
Since the Schr\"odinger potential is positive semi-definite, we can
immediately conclude that $\chi_{\rm a}=0$ and this in turn tells us
from the second equation of (\ref{eq:fluc_equations_YM}) that also
$\chi_{\rm f}$ needs to vanish on the entire plane. It
follows that there are no normalizable massless modes in the Abelian
system. 

For the non-Abelian case, the Schr\"odinger equation seems to be
somewhat more complicated.  
There is however a nice trick to solve this problem, which is to write
the square of the adjoint operator on a given state as a sum of
squared terms. It was first introduced in Ref.~\cite{Hanany:2003hp} in
the case of $G'=SU(N)$, and was also applied to a general group in
Ref.~\cite{Eto:2009bg}. 
The starting point is to take the complex norm of 
$\mathbb{D}^\dag\chi = 0$ as
\beq
0 &=& \left({\mathbb{D}}^\dag \chi, {\mathbb{D}}^\dag \chi\right) \non
&=& \int_{\mathbb{C}}\Tr\left[
\left|i\D_{\rm f} \chi_{\rm f} + \chi_{\rm a}H\right|^2
+\left|i\bar\D_{\rm a} \chi_{\rm a} - 
  2\left<\chi_{\rm f} H^{\dag}\right>_G\right|^2
\right] \non
&=& \int_{\mathbb{C}}\Tr \left[
\left|\D_{\rm f}\chi_{\rm f}\right|^2 + 
\left|\bar\D_{\rm a}\chi_{\rm a}\right|^2 + 
\left|\chi_{\rm a} H\right|^2 + 
4\left|\left<\chi_{\rm f}H^\dag\right>_G\right|^2 + c
\right] \ ,
\label{eq:trick_HT}
\eeq
where we have set $e^2=4$ and the cross term $c$ has been rewritten
into two total derivative terms by using the BPS equation 
$\bar{\D}H = 0$:
$
c = \Tr\left[i\p\left(\chi_{\rm f}H^\dag \chi_{\rm a}^\dag\right) 
-i\bar\p\left(\chi_{\rm a} H \chi_{\rm f}^\dag\right)\right]
$.
Since we are interested in normalizable physical zero-modes, we
require the fluctuations to vanish at spatial infinity
($|z|\to\infty$). Thus the total derivative terms vanish. 
We now want to show
$\chi_a = 0$ and $\chi_f = 0$ from the conditions
\beq
\D_{\rm f}\chi_{\rm f} = 0 \ , \quad
\bar\D_{\rm a}\chi_{\rm a} = 0 \ , \quad
\chi_{\rm a} H = 0 \ , \quad
\left<\chi_{\rm f}H^\dag\right>_G = 0 \ .
\eeq
The $N$-by-$\NF$ matrix field $H$ (background field) 
has the maximum rank $N$ almost everywhere 
except at 
the vortex centers. Thus we conclude $\chi_{\rm a}=0$ from
$\chi_{\rm a}H=0$.
For $\chi_{\rm f}$ we need a more elaborated argument except for 
the case of $G=U(N)$ and $\NF=N$ \cite{Hanany:2003hp}.
Let us explain $\chi_{\rm f}=0$ for the case of $G'=SU(N)$ and $\NF >
N$.\footnote{We thank K.~Ohashi for this point.}
First, we solve $\bar\D \chi_{\rm f}^\dag = 0$ by 
a combination of the background field $S$ and a holomophic matrix function
$\tilde{H}_0(z)$ by $\chi_{\rm f}^\dag = \tilde H_0(z) S$.
Then consider a holomorphic $G'=SU(N)$ invariant operator
(i.e.~an $\NF$-by-$\NF$ matrix) 
$M \equiv \chi_{\rm f}^\dag H = \tilde{H}_0(z) H_0(z)$.
Note that
$\chi_{\rm f}^\dag$ can be reconstructed from $M$,
because $H$ has the maximal rank in the broken (Higgs) vacuum.
However, $M$ vanishes at infinity because $\chi_{\rm f}$ should be
normalizable, hence we require $\tilde{H}_0(z) H_0(z) \to 0$. The
holomophic matrix $\tilde{H}_0(z) = 0$ is a unique solution. 
Thus we conclude $\chi_{\rm f} = 0$.
For the generic simple group $G'$, similar arguments were done in
Ref.~\cite{Eto:2009bg}.

Now we have confirmed that ${\mathbb D}^\dag$ has no normalizable
zero-modes. So the index given in Eq.~(\ref{eq:res_index_ym}) indeed
counts the correct number of the physical zero-modes.


\subsection{Chern-Simons-Higgs theory}


\subsubsection{The index for Chern-Simons vortices}

Next we study the index of the BPS system in the non-Abelian
CS theory. For simplicity, we set $\kappa = \mu$, hence we
will be using Eqs.~(\ref{eq:NACSmastersystem1}) and
(\ref{eq:CSequalcouplingBPS}). 
To the best of our knowledge, the index theorem for the CS 
vortices has been made only in the Abelian case in
Refs.~\cite{Jackiw:1990pr,Lee:1991td}. With the formula
(\ref{eq:effectivematrix}), we can easily calculate the index also in
the non-Abelian case with a generic gauge group $G'$.  
The fluctuations of the BPS equations read to linear order
\begin{align}
i\bar{\D}\,\delta\!H - \delta\!\bar{A}H &= 0 \ , 
\label{eq:CSfluc1} \\ 
i\bar{\D}\,\delta\!A - i\D\,\delta\!\bar{A} &=
\frac{16\pi^2}{\kappa^2}
\left<
\left<\delta\!H H^\dag\right>_G H H^\dag
+ \left<H H^\dag\right>_G \delta\!H H^\dag - \frac{v^2}{2}\delta\!H H^\dag
\right>_G \non
& + \frac{16\pi^2}{\kappa^2}
\left<
\left<H \delta\!H^\dag\right>_G H H^\dag
+ \left<H H^\dag\right>_G H \delta\!H^\dag - \frac{v^2}{2}H\delta\!H^\dag
\right>_G \ . \label{eq:CSfluc2}
\end{align}
We choose the gauge fixing condition as
\begin{align}
i\bar{\D}\,\delta\!A + i\D\,\delta\!\bar{A} =
&-\frac{16\pi^2}{\kappa^2}
\left<
\left<\delta\!H H^\dag\right>_G H H^\dag
+ \left<H H^\dag\right>_G \delta\!H H^\dag - \frac{v^2}{2}\delta\!H H^\dag
\right>_G \non
& + 
\frac{16\pi^2}{\kappa^2}
\left<
\left<H\delta\!H^\dag\right>_G H H^\dag
+ \left<H H^\dag\right>_G H\delta\!H^\dag - \frac{v^2}{2}H\delta\!H^\dag
\right>_G \ . \label{eq:CSfluc3}
\end{align}
Combining Eqs.~(\ref{eq:CSfluc2}) and (\ref{eq:CSfluc3}), we obtain
\beq
i\D\,\delta\!\bar{A} =
- \frac{16\pi^2}{\kappa^2}
\left<
\left<\delta\!H H^\dag\right>_G H H^\dag
+ \left<H H^\dag\right>_G \delta\!H H^\dag - \frac{v^2}{2}\delta\!H H^\dag
\right>_G \ . \label{eq:CSfluc4}
\eeq

Taking the transpose of the Eqs.~(\ref{eq:CSfluc1}) and
(\ref{eq:CSfluc4}), we can now rewrite the above linear equations as 
the Dirac-type equation (\ref{eq:dirac}) with the operator
$\mathbb{D}$ defined as
\beq
\mathbb{D} = 
\begin{pmatrix}
i\bar{\D}_{\rm f}^{\rm T} & -H^{\rm T} \\
\frac{16\pi^2}{\kappa^2}
\left<
\bar{H} H^{\rm T} \left<\bar{H}\circ\right>_G 
  +\bar{H} \circ\left<\bar{H} H^{\rm T}\right>_G  
  -\frac{v^2}{2}\bar{H} \circ
\right>_G
 & i\D^{\rm T}_{\rm a}
\end{pmatrix} \ .
\eeq
Although this Dirac operator takes a somewhat complicated form, it is
indeed in the class of operators (\ref{eq:operatorclass}). Thus we can
immediately apply the result of Sec.~\ref{sec:indexcalc} also to this
case and hence the index is given by Eq.~(\ref{eq:index}).

Let us explain the result for each type of vacuum.
First for the topological solitons
the result is very sound and the index is exactly
\beq 
\mathcal{I} = \hat{\mathcal{I}} = \NF N \nu = \frac{\NF N k}{n_0}\ .
\label{eq:index_topological}
\eeq
Note that the index is always an integer and this result depends on
the gauge group $G'$ only through $n_0$. 
We will see that this index indeed counts the number of zero-modes.

In the case of the non-topological solitons in the unbroken vacuum,
the magnetic flux is not quantized, so $\hat{\cal I}$ is not an
integer
\beq
\hat {\cal I} = \frac{ \NF N (k' + \alpha)}{n_0} \ ,
\eeq
with $k' \in \mathbb{Z}_{\ge 0}$ and $\alpha\in \mathbb{R}_{+}$.
However, since the index ${\cal I}$ counts the number of the
zero-modes, it should give an integer.
In order to obtain the correct result, we have to subtract
contributions from the continuum due to unbounded resonances
\cite{Kiskis:1977vh}. Thus the correct index is given by
\beq
{\cal I} = \frac{ \NF N (k' + \hat\alpha)}{n_0} \ ,
\label{eq:index_cs_nt}
\eeq
where $\hat\alpha$ is the largest integer less than $\alpha$
($\alpha-\hat{\alpha}<1$).
$k'$ is the number of topological solitons living ``inside'' the
non-topological solitons and the number is limited by the amount of
non-topological winding $\alpha$. 
Note that this index still includes some non-physical zero-modes which 
change the magnetic flux \cite{Jackiw:1990pr}. We will return this point later.

When the topological and non-topological solitons coexist, the index
is mixed as
\beq
{\cal I} = \frac{ \NF N k}{n_0} + \frac{ \NF N (k' + \hat\alpha)}{n_0},
\eeq
where the first term is related to the topological solitons while
the second term is related to the non-topological solitons.
Here we intend that the $k$ vortices are topological in the
(partially) broken vacuum, while the $k'$ vortices are topological
vortices living ``inside'' the non-topological vortices in the
(partially) unbroken vacuum.


\subsubsection{Vanishing theorem for Chern-Simons vortices\label{sec:vanish_CS}}

Next, we would like to make a vanishing theorem
(\ref{eq:vanishing_th}) for $\mathbb{D}^\dag$ in the CS
case. The adjoint operator reads
\beq
\mathbb{D}^\dag =
\begin{pmatrix}
i\D_{\rm f}^{\rm T} & 
\frac{8\pi^2}{\kappa^2} H^{\rm T} \left(
\left<\bar{H}H^{\rm T}\circ\right>_G 
+ \circ\left<\bar{H}H^{\rm T}\right>_G
- \frac{v^2}{2} \right) \\
- 2\left<\bar{H}\circ\right>_G & i\bar\D^{\rm T}_{\rm a}
\end{pmatrix} \ .
\label{eq:adj_cs}
\eeq
The question is if there exist normalizable zero-modes $\chi$ 
\beq
\mathbb{D}^\dag \chi = 0 \ , \qquad
\chi = 
\begin{pmatrix}
\chi_{\rm f}^{\rm T}\\
\chi_{\rm a}^{\rm T}
\end{pmatrix} \ ,
\eeq
where $\chi_{\rm f}^{\rm T}$ is an $\NF$-by-$N$ matrix and 
$\chi_{\rm a}^{\rm T}$ is an $N$-by-$N$ matrix.
Unfortunately, we cannot use the same trick as we used in
Eq.~(\ref{eq:trick_HT}) since the cross terms associated with the
completion of the square do not vanish even by the use of the BPS
equations.  
Therefore, we have to solve the problem in a more straightforward way. 
Let us write down the Dirac equation as follows
\begin{align}
\D_{\rm f} \chi_{\rm f} &= \frac{i8\pi^2}{\kappa^2} 
\left(
\left<\chi_{\rm a} H H^\dag \right>_G 
+ \left<H H^\dag\right>_G \chi_{\rm a}
- \frac{v^2}{2} \chi_{\rm a}
\right) H \ ,  
\label{eq:CSvanishingeq1} \\ 
\bar{\D}_{\rm a} \chi_{\rm a} &= -i2\left<\chi_{\rm f} H^\dag\right>_G
  \ ,  \label{eq:CSvanishingeq2}
\end{align}
Taking the adjoint derivative $\D_{\rm a}$ of the second equation, and
using the BPS equations, allows us to combine the two equations as
\begin{align}
\D_{\rm a}\bar{\D}_{\rm a}\chi_{\rm a} 
= \frac{16\pi^2}{\kappa^2}
\left<\left(
\left<\chi_{\rm a} H H^\dag \right>_G 
+ \left<H H^\dag\right>_G \chi_{\rm a}
- \frac{v^2}{2} \chi_{\rm a}
\right) H H^\dag\right>_G \ . 
\label{eq:CSvanishingeqcombined}
\end{align}
This is a non-Abelian Schr\"odinger equation for $\chi_{\rm a}$ with
the background configuration $A_i$ and $H$. 
This tells us that $\mathbb{D}^\dag$ has a non-zero kernel if and only
if the Schr\"odinger equation has a normalizable zero-eigenstate. 

Unfortunately, up to now, no proofs have been given even in the simplest Abelian case
but $\chi_{\rm a}=0$ has been expected in general 
not to have normalizable zero-modes~\cite{Jackiw:1990pr,Lee:1991td}.
For the non-Abelian cases, we have nothing 
to add beyond the Abelian case. We would in general expect
that $\chi_{\rm a} = 0$ and $\chi_{\rm f}=0$ holds
and that the index (\ref{eq:index}) counts the complex dimension of the
kernel of $\mathbb{D}$ and hence the number of bosonic zero-modes of
the BPS configuration.
A partial evidence for us to expect this is that we will explicitly realize
the same number of the physical zero-modes for the topological solitons 
as is given by the index theorem result (\ref{eq:index_topological}).


\section{The moduli matrix method\label{sec:modulimatrix}}

In Sec.~\ref{sec:indextheorem}, we have demonstrated the existence of
a certain number of zero-modes for the BPS solutions by using a formal 
mathematical computation of the index of a Dirac operator describing
the BPS configurations under consideration. 
Here in this section, on the contrary, we will realize the zero-modes
in a more explicit manner. The so-called moduli matrix method 
\cite{Isozumi:2004jc,Isozumi:2004vg,Eto:2005yh,Eto:2006pg} is
suitable for that purpose which was first developed to describe the
topological solitons in the Higgs phase of supersymmetric YM
theories without CS terms. 

Let us apply the moduli matrix for the vortex system
\cite{Eto:2005yh,Eto:2006pg,mmv} in order to solve the first of the
BPS equations (\ref{eq:YMBPS1}) and (\ref{eq:NACSmastersystem1})
(which is the same in both the YM and CS theories): 
\beq
H = {S}^{-1}H_0(z) \ , \qquad 
\bar{A} = -iS^{-1}\bar{\p}S \ , \qquad
S \in {G}^{\mathbb{C}} \ ,
\label{eq:mm}
\eeq
where $H_0(z)$ is an $N$-by-$\NF$ complex matrix which is holomorphic 
in $z$. $H_0(z)$ is called the moduli matrix
\cite{Isozumi:2004jc,Isozumi:2004vg,Eto:2005yh,Eto:2006pg} 
and the decomposition of the gauge field is called the Karabali-Nair
form \cite{Karabali:1995ps}. 
$S$ takes a value in the
complexified gauge group $G^{\mathbb{C}}$.
For instance for $G=U(N)$ the complexified gauge symmetry is
$GL(N,\mathbb{C})$. 
Note that the one-form $\bar{A}$ is not pure gauge. 
The gauge symmetry and the flavor symmetry act on the new matrices as 
follows 
\beq
\left(S^{-1},\ H_0\right) \to
  \left(U_{\rm c} S^{-1},\ H_0 U_{\rm f}\right) \ .
\eeq
Furthermore, there is an equivalence relation, which is
denoted $V$-equivalence and it acts as
\beq
\left(S,\ H_0\right) \sim V(z) \left(S,\ H_0\right) \ , \quad
V(z) \in {G}^{\mathbb{C}} \ , 
\label{eq:v-equiv}
\eeq
with all the elements in $V(z)$ being any holomorphic function with
respect to $z$. In order not to change the winding number (energy) of
the solution, we should impose that the determinant of $V(z)$ is
non-zero.

That is all for the first equation. 
The moduli matrix $H_0(z)$ has all the information about the
topological solitons in the YM theories
\cite{Isozumi:2004vg,Eto:2005yh}.   
It is conjectured in Ref.~\cite{Gudnason:2009ut} that the same holds
for the topological solitons in the (non-Abelian) CS theories. 

Since the classification of the moduli space for the non-Abelian
vortices has been carried out in the case of YM theories, we
will not repeat it here. In the case of CS theories, the
results of Ref.~\cite{Gudnason:2009ut} claim that the moduli spaces
of the YM vortices apply also to the Higgs phase of the
CS theories. However, in the case of CS theory --
as already mentioned in Sec.~\ref{sec:model_cs} -- there exist not
only the Higgs phase but also the CS phase (unbroken phase)
and the partially broken phases. In these cases, we cannot use the
arguments of Ref.~\cite{Gudnason:2009ut} and furthermore, as we will
see later, the moduli matrix is not sufficient to describe all of the
moduli parameters possessed by the vortices. 

After introducing the $S$ field in Eq.~(\ref{eq:mm}), it is 
natural to introduce a local gauge invariant field
\cite{Isozumi:2004vg,Eto:2005yh,Eto:2006pg} 
\beq 
\Omega = S S^\dag \ , \quad \Omega = \omega\hat{\Omega} \ . 
\eeq 
Note that the complex matrix field $S$ can be decomposed as 
$S = s\hat{S}$ with the Abelian part 
$s\in U(1)^{\mathbb{C}}\sim \mathbb{C}^\star$ and the
non-Abelian part $\hat{S}\in{G'}^{\mathbb{C}}$. 
In the right equation above, we have defined $\omega = |s|^2$ and
$\hat{\Omega} = \hat{S}\hat{S}^\dag$.
We will see that this $\Omega$ will play a central role in the
following sections. 
Note that $\Omega$ transforms as $\Omega \to V \Omega V^\dag$ with
respect to the $V$-transformation. 
Let us also introduce a ``current'' 
defined as \cite{Karabali:1995ps}
\beq
{\cal J} \equiv \p \Omega \Omega^{-1} \ .
\eeq
This ${\cal J}$ transforms as a holomorphic connection under the
$V$-transformation 
\beq
{\cal J} \to V {\cal J} V^{-1} + \p V V^{-1} \ .
\eeq
In terms of ${\cal J}$, the magnetic field can be rewritten as 
\beq F_{12} = - 2 S^{-1} \bar\p J S \ ,
\eeq
(with $F^0_{12} = -2\sqrt{2N}\bar{\p}\p\log\omega$ and 
$\hat{F}_{12}= -2\hat{S}^{-1}\bar{\p}(\p\hat{\Omega}\hat{\Omega}^{-1})\hat{S}$). 
The magnetic flux is expressed as
\beq
\Phi = - \int_{\mathbb{C}} \Tr [F_{12}] 
= 2 \Tr\int_{\mathbb{C}} \bar\p {\cal J}
= 2N \int_{\mathbb{C}} \bar{\p}\p\log\omega \ ,
\eeq
while the $U(1)$ winding number is given by
\beq 
\nu = \frac{1}{\pi} \int_{\mathbb{C}} \bar{\p}\p\log\omega \ .
\eeq
Let us define a covariant derivative with the connection $i {\cal J}$.
For example, for an adjoint field $\phi$, transforming as 
$\phi\to V\phi V^{-1}$, then we have
\beq
\D_v \phi = \p\phi - [{\cal J},\phi] \ .
\label{eq:coDv}
\eeq
$\D_v\phi$ transforms homogeneously under the $V$-transformation.


\subsection{Yang-Mills-Higgs theory \label{sec:mm_ym}}

For the readers who are not familiar with the moduli matrix, let us
review several examples (the details of this topic can be found in
Ref.~\cite{Eto:2006pg}). 
As the first example, let us review the Abelian vortex. The moduli
matrix for $k$ vortices reads
\beq
H_0(z)= a_0 + a_1 z + a_2 z^2 + \cdots + a_{k-1}z^{k-1} + z^k
= \prod_{i=1}^k(z-z_i) \ ,
\eeq
where we have fixed the coefficient of $z^k$ to unity by using the
$V$-equivalence relation. 
The set of complex parameters $\{a_0,a_1,\ldots, a_{k-1}\}$ are the
free parameters (the position moduli) of the BPS solutions. Hence, the
moduli space of the $k$ Abelian vortices is 
\beq
{\cal M}^k_{U(1)} = \mathbb{C}^k \ .
\eeq
The complex dimension is $k$, which is consistent with the index
theorem result (\ref{eq:index}). 

The second example we will review here is the $k=1$ vortex in the
$G'=SU(N)$ model with $\NF=N$ flavors. After fixing the
$V$-equivalence, one can obtain the following matrix 
\beq
H_0(z) = 
\begin{pmatrix}
1 & & & b_1\\
& \ddots & & \vdots\\
& & 1 & b_{N-1} \\
& & & z - z_1
\end{pmatrix} \ .
\eeq
The set of complex parameters $\{b_1,\cdots,b_{N-1}\}$ corresponds to 
the inhomogeneous coordinate of the orientational zero-modes
$\mathbb{C}P^{N-1}$ while $z_0$ is the position (translational)
zero-mode. Hence, we obtain the corresponding moduli space
\beq
\mathcal{M}_{U(N)}^{k=1} = \mathbb{C} \times \mathbb{C}P^{N-1} \ .
\eeq
The complex dimension is $N$ which is again consistent with the index 
theorem (\ref{eq:index}). In this way, the moduli matrix provides us
a realization of the physical zero-modes. 

We would like to emphasize that the use of the moduli matrix is quite
easy. One just needs to choose a holomorphic matrix which satisfies
the conditions posed on the invariants with respect to $G'$ and
it should be consistent with the chosen boundary conditions. 
For instance, any matrix $H_0(z)$ whose determinant is a polynomial in 
$z$ of degree $k$, generates $k$ non-Abelian vortices in the case of 
$G'=SU(N)$. 
When we choose the form of the moduli matrix $H_0(z)$ we should fix
the $V$-equivalence in order to eliminate unphysical complex
parameters.  
Once we have done this, one can easily read off the physical moduli
parameters for the solutions. 
For a generic gauge group $G'$, we have to take further conditions
into account. 
We will not discuss those details here as they can be found in
Refs.~\cite{Eto:2008yi,Eto:2009bg}.

In order to solve completely the BPS equations, we now have to solve 
the second equation of the BPS equations (\ref{eq:YMBPS1}). 
Unfortunately, this is not an easy task since no analytic solutions
have been obtained even in the case of the simplest Abelian models. 
Solving the second equation is very important in the following two
senses: 
\begin{enumerate}
\item [1)] We have to confirm the existence of the BPS solution for
  each given moduli matrix. 
  If the solution does not exist, it means that the complex parameters
  in the moduli matrix cannot be moduli parameters. 
\item [2)] We have to confirm uniqueness of the solutions of the second
  equation. If we are not able to show the uniqueness, it means that
  there might exist additional zero-modes which are not included in
  the moduli matrix description. 
\end{enumerate}
These problems have been resolved only in the case of the $k$ vortices
in the Abelian-Higgs model by Taubes \cite{Taubes:1979tm}. 
But there is no available proof for the non-Abelian
cases\footnote{This is true on the infinite $\mathbb{C}$-plane. See
  however Ref.~\cite{Baptista:2008ex} for non-Abelian $U(N)$ vortices
  on compact Riemann surfaces. }. 

The strongest statement in favor of our belief of existence of the
solutions in the non-Abelian models is that several numerical
solutions have been found explicitly in the literature. Hence, we are
quite confident that the existence holds in all our models, although
we currently do not have a rigorous mathematical proof.

On the uniqueness problem, we have observed that the number of the
physical complex parameters in the moduli matrix coincides with that 
of the index theorem (\ref{eq:index}) in all known models
\cite{Hanany:2003hp,Eto:2009bg}.
Although this coincidence is in favor of our intuition, we still miss
a direct relation between the uniqueness problem and the index
theorem.
We will partially solve the uniqueness problem of the non-Abelian
vortices in YM and CS theories in
Sec.~\ref{sec:uniqueness}.  

Let us rewrite the second equation of (\ref{eq:YMBPS1})
in terms of $\Omega$ and the moduli matrix $H_0(z)$
\begin{align}
\bar{\p}\p\log\omega &=
\frac{e^2}{4N}\Tr\left[v^2{\bf 1}_N - 
  \Omega_0\Omega^{-1}\right]
  \ , \label{eq:YMmasterA}\\ 
\bar{\p}(\p\hat{\Omega}\hat{\Omega}^{-1}) &= 
  -\frac{g^2}{2}\left<\Omega_0\Omega^{-1}\right>_{G'} \ , 
  \label{eq:YMmasterNA}
\end{align}
with $\Omega_0\equiv H_0 H_0^\dag$ and $\Omega=\omega\hat{\Omega}$. 
We have used the following relation
\beq
\left<X\right>_{G'} = \Tr\left[X t^a\right] t^a 
= \Tr\left[X \tilde{t}^a\right] \tilde{t}^a \ , \quad
\tilde{t}^a \equiv \hat{S} t^a \hat{S}^{-1} = S t^a S^{-1} \ , 
\eeq
where $\tilde{t}^a$ is an automorphism of $t^a$. 
These equations are called the master equations for the YM
vortices. 
The equal coupling case, $g=e$ can for a generic group neatly be
written as 
\beq
\bar\p(\p\Omega\Omega^{-1})
= \frac{e^2}{2} \left<v^2{\bf 1} - \Omega_0\Omega^{-1} \right>_G \ .
\label{eq:masterYM_equal}
\eeq

Before closing this section, let us see the asymptotic behavior of 
$\Omega$. It is again sufficient to consider the minimal winding
solution in the Abelian model. 
Hence we take $H_0(z) = z$ and then $\omega$ approaches 
$\Omega_0/v^2 = |z|^2/v^2$. 
Plugging this into the master equation, one easily obtains the
asymptotic solution 
\beq
\omega = \frac{|z|^2}{v^2}\left[1 + q K_0(m_e |z|)\right] \ ,
\label{eq:asym_ah}
\eeq
where $K_0$ is the modified Bessel function of the second kind and
$q$ is an unknown constant parameter which can be determined
numerically.

\subsection{Chern-Simons-Higgs theory}

As mentioned above, the first BPS equation is common for the
YM models and the CS models. Hence, the moduli
matrix method which solves the first BPS equation explained in
Sec.~\ref{sec:mm_ym} can be applied to the CS models without
any modifications \cite{Gudnason:2009ut}. 
This is one of the significant features of the moduli matrix method,
namely its application range is indeed quite large.

Since we have already solved the first equation, the remaining task 
is to solve the second equation of (\ref{eq:NACSmastersystem1}).
In Ref.~\cite{Gudnason:2009ut}, the BPS equations were given for a
generic gauge group $G$, but the master equations were provided only
for $G'=SU(N),SO(N),USp(2M)$. Here we will provide the master
equations completely independent of the choice of $G'$
\begin{align}
\bar{\p}\p\log\omega &= 
\frac{4\pi^2}{N^2\kappa^2}
  \Tr\left[v^2{\bf 1}_N - \Omega_0\Omega^{-1}\right]
  \Tr\left[\Omega_0\Omega^{-1}\right]
-\frac{8\pi^2}{N\kappa\mu}
  \Tr\left[\left<\Omega_0\Omega^{-1}\right>_{G'}\Omega_0\Omega^{-1}\right] \ ,
  \label{eq:CSmasterA}
\\
\bar\p(\p\hat\Omega\hat\Omega^{-1}) &=
\frac{8\pi^2}{N\kappa\mu}
  \Tr\left[v^2{\bf 1}_N - \Omega_0\Omega^{-1}\right]
  \left<\Omega_0\Omega^{-1}\right>_{\!\!G'}
-\frac{16\pi^2}{\mu^2}
  \left<\left<\Omega_0\Omega^{-1}\right>_{\!\!G'}
  \Omega_0\Omega^{-1}\right>_{\!\!G'} \ .
\label{eq:CSmasterNA}
\end{align}
The equal coupling case, $\mu=\kappa$ can neatly be written for a
generic group as
\beq
\bar{\p}(\p\Omega\Omega^{-1}) = \frac{16\pi^2}{\kappa^2}
\left<\left<v^2{\bf 1}_N - \Omega_0\Omega^{-1}\right>_G
  \Omega_0\Omega^{-1}\right>_G \ .
\label{eq:masterCS_equal}
\eeq
These are called the master equations for the CS theory. 

Concerning the master equation, we have the same problems as in the
YM theories. Namely, the existence and uniqueness of the
solutions. 
To the former problem, we are in the same situation as in the case of
YM theories, the best argument in favor of the existence we
have currently, is the various numerical solutions to the non-Abelian
CS BPS equations \cite{Gudnason:2009ut}.
The existence of the solutions however is an important future problem
but it is beyond the scope of this paper. In the Abelian case, the
existence 
however has been proved in the topological case
\cite{Wang:1991na} and in the non-topological case
(for radially symmetric solutions)
\cite{Spruck:1992yy}. 

As for the uniqueness problem, we will partially solve it in the
Sec.~\ref{sec:uniqueness}. 
But there is a big difference between the YM theories and the 
CS theories.
As we will see in Sec.~\ref{sec:uniqueness}, the solutions to the
master equation have their own moduli parameters, when we choose a vacuum
different from the Higgs vacuum at the boundary.

\subsubsection{Abelian Chern-Simons solitons}

We will now briefly describe the Abelian solutions
\cite{Jackiw:1990pr} in the moduli matrix formalism with $\NF=1$
flavor. 
Starting with the vacuum configurations. There are two vacua: i) the
broken vacuum and ii) the unbroken vacuum. 
The Higgs field is $H=v$ (up to $U(1)$ gauge symmetry) in the former
case, so it is described by
\beq
\text{broken vacuum}:\ 
H_0 = 1 \ , \quad s^{-1} = v \ .
\eeq
Note that the $V$-equivalence has been fixed by the choice $H_0=1$. 
In the unbroken vacuum, $H=0$, we choose the moduli matrix as follows
\beq
\text{unbroken vacuum}:\ 
H_0 = 0 \ , \quad s^{-1} = 1 \ ,
\eeq
Here we have set $s = 1$ by using the $V$-equivalence relation.
At first glance, the $U(1)$ gauge symmetry seems to be broken since 
$s^{-1}$ transforms as $s^{-1} \to g s^{-1}$ with $g\in U(1)$.
But this transformation can be absorbed by an according
$V$-transformation, such that no symmetries are broken. 

Let us next consider $k$ topological vortices. It is generated by
the moduli matrix 
$H_0 = \prod_{i=1}^k(z-z_i)$. 
The number of moduli parameters is $k$ and is in accord with the index (\ref{eq:index}). 
The master equation determines $\omega$ as
\beq
\bar{\p}\p\log\omega = \frac{4\pi^2}{\kappa^2} \prod_{i=1}^k|z-z_i|^{2}\omega^{-1}
\left(v^2 - \prod_{i=1}^k|z-z_i|^{2}\omega^{-1}\right) \ .
\label{eq:mm_k1_tpcs}
\eeq
We impose the boundary condition for $\omega$ in such a way that
the Higgs field approaches to the Higgs phase $|H| \to v$.
With respect to $\omega$, this is equal to imposing
the boundary condition
\beq
\omega \to v^{-2}\prod_{i=1}^k|z-z_i|^{2},\qquad \text{as} \quad |z| \to \infty.
\eeq
Note that this boundary condition is unique for obtaining a regular
solution. 
We would like to stress that all the moduli parameters of the
topological vortices are included in the moduli matrix. 
Unfortunately, no analytic solutions to this equation even for $k=1$
are known. The asymptotic solution of $k=1$ however is 
\beq
\omega_{k=1} = v^{-2}|z|^{2}
\left[1 + \tilde{q} K_0\left(m_{\kappa\infty} |z|\right)\right] \ ,
\label{eq:asym_cs_top}
\eeq
where $\tilde{q}$ is an unknown parameter which can be determined
numerically. There is no difference in the asymptotic form of the
topological solitons of the Abelian-Higgs model and the CS
model, see Eqs.~(\ref{eq:asym_ah}) and (\ref{eq:asym_cs_top}). 
This is because they are topological solitons in the broken
vacuum. A tiny difference lies in the masses of the Higgs fields $m_e$ 
versus $m_{\kappa\infty}$ and in the numerical coefficients $q$ versus 
$\tilde{q}$. 

Let us next consider the non-topological vortex with $k$ Higgs zeros 
and magnetic flux 
$\Phi=-\int_{\mathbb{C}} F_{12}=2\pi(k+\alpha)$.
The total magnetic flux is not necessarily an integer. In fact,
$\alpha$ can be an arbitrary real number. 
We again take the moduli matrix $H_0 = \prod_{i=1}^k(z-z_i)$.
Then the master equation is also the same as Eq.~(\ref{eq:mm_k1_tpcs}).
We solve this with a different boundary condition for $\omega$ 
consistent with $H \to 0$.
It is determined by the total amount of magnetic flux 
\beq
F_{12} = -2\bar{\p}\p\log\omega \to -2\bar{\p}\p\log |z|^{2(k+\alpha)}
\ ,
\eeq
for $|z|\to\infty$. Thus the desired boundary condition reads
\beq
\omega \to C^{-1} |z|^{2\alpha} \prod_{i=1}^k|z-z_i|^2  \ ,
\label{eq:bc_nt_u1}
\eeq
where $C$ is a unique numerical constant.
$H$ asymptotically approaches the unbroken vacuum as 
$H \to C/|z|^{\alpha}$. 
In this way, we have found $k$ zero-modes $\{z_i\}$ in the moduli
matrix as in the case of topological vortices. However, 
in Sec.~\ref{sec:unique_cs}, we will see that Eq.~(\ref{eq:bc_nt_u1})
is not the most generic boundary condition on $\omega$ for the soliton
solutions with a fixed magnetic flux $\Phi=2\pi(k+\alpha)$. 
We will also see that there exist additional zero-modes which are not 
accounted for in the moduli matrix but on the other hand reside in
$\omega$.

\subsubsection{$U(2)$ Chern-Simons solitons}

Let us next explain the solitons in the $U(2)$ CS theory
($\kappa=\mu$) with $\NF=2$ in terms of the moduli matrix.

We start by describing the vacua as we did in the Abelian case above. 
As we have explained in Sec.~\ref{sec:u2_cs}, there are three vacua in 
the case at hand: (2) the fully broken vacuum, (1) the partially broken
vacuum and (0) the unbroken vacuum. 
The Higgs fields corresponding to these vacua are given in
Eq.~(\ref{eq:vac_u2_cs}), while the corresponding moduli matrices are
given by 
\beq
H_0^{(2)} = v
\begin{pmatrix}
1 & 0 \\
0 & 1
\end{pmatrix} \ , \qquad
H_0^{(1)} = v
\begin{pmatrix}
\alpha & \beta \\
0 & 0
\end{pmatrix} \ , \qquad
H_0^{(0)} = 
\begin{pmatrix}
0 & 0 \\
0 & 0
\end{pmatrix} \ .
\eeq
Here $\alpha,\beta\in\mathbb{C}$ parametrize the vacuum moduli space
i.e.~$\mathbb{C}P^1$. 
We fix these parameters as $(\alpha,\beta)=(1,0)$ by using
$V$-equivalence and flavor symmetry. 
For all the vacuum states, $S$ is chosen to be the unit matrix.

Let us next consider the solitons.
Since we are working with $\kappa=\mu$, most solutions in the $U(2)$
case can be obtained by the embedding of some Abelian solutions.
For instance, the $k=1$ topological vortex can be obtained by
\beq
H_0 = \begin{pmatrix}
z-z_0 & 0 \\
b & 1
\end{pmatrix} \ , \qquad
\Omega = 
\begin{pmatrix}
\frac{\omega_{\rm s} + |b|^2|z-z_0|^2}{1+|b|^2} & b^* (z-z_0) \\
b (z-z_0)^* & 1 + |b|^2
\end{pmatrix} \ ,
\eeq
where $\omega_{\rm s}$ is the $k=1$ solution of the Abelian master equation 
(\ref{eq:mm_k1_tpcs}) with the boundary condition 
$\omega_{\rm s} \to v^{-2}|z|^2$.
The $U(1)$ winding number is not integer but half-integer
$\nu=\frac{1}{\pi}\int_{\mathbb{C}}\bar{\p}\p\log\omega_{\rm s}^{\frac{1}{2}}=1/2$.
We have realized two complex moduli parameters 
$z_0$ and $b$ in the moduli matrix. The former is the position modulus
and the latter is the orientational modulus parameter of
$\mathbb{C}P^1$, see Ref.~\cite{Eto:2004rz} for details. 
Hence, we have two moduli ($\NF N \nu = 2$) which is in accord with
the index result (\ref{eq:index}). 

Next we study solitons in the partially broken vacuum.
We are interested in the semi-local vortex in this vacuum.
The moduli matrix and $\Omega$ are respectively given by
\beq
H_0 = 
\begin{pmatrix}
z-z_0 & a \\
0 & 0
\end{pmatrix} \ , \qquad
\Omega = 
\begin{pmatrix}
\omega_{\rm s} & 0 \\
0 & 1
\end{pmatrix} \ ,
\eeq
where $z_0$ again denotes the position. 
The complex parameter $a$ is a combination of the transverse size
$|a|$ and a relative phase ${\rm Arg}~a$, while $\omega_{\rm s}$ is
the solution to the master equation 
\beq
\bar{\p}\p\log\omega_{\rm s} = \frac{4\pi^2}{\kappa^2}
\left[v^2 - \frac{|z-z_0|^2+|a|^2}{\omega_{\rm s}}\right]
\frac{|z-z_0|^2+|a|^2}{\omega_{\rm s}} \ .
\eeq
This equation should be solved with the boundary condition
\beq
\omega_{\rm s} \to \frac{|z-z_0|^2 + |a|^2}{v^2} \ .
\eeq
Thus, the $U(1)$ winding number is again a half integer
$\nu = \frac{1}{\pi}\int_{\mathbb{C}} \p\bar\p\log
\omega_{\rm s}^{\frac{1}{2}} = 1/2$.
Hence, we find again two moduli parameters in accord with the index
result (\ref{eq:index}). 
In the Abelian-Higgs model, the zero-mode $a$ of the semi-local vortex
is a non-normalizable zero-mode. 
We suspect that $a$ in the semi-local CS soliton is also
non-normalizable. 

One can also consider the semi-local non-topological solitons in the
partially unbroken vacuum. 
We do not work it out in this paper, although it might be interesting. 
The reader who is interested in this topic can easily extend our
moduli matrix method to this case by comparing our method and the
results in Ref.~\cite{Khare:1992qr}.

Finally, we study the non-topological solitons in the fully unbroken
vacuum. 
To this end, we choose the following moduli matrix and $\Omega$ 
\beq
H_0 = 
\begin{pmatrix}
1 & b \\
0 & 0
\end{pmatrix} \ , \qquad
\Omega = 
\begin{pmatrix}
(1+|b|^2)\omega_{\rm s} & 0 \\
0 & 1
\end{pmatrix} \ ,
\label{eq:mm_cs_nt_ori}
\eeq
where $\omega_{\rm s}$ satisfies the master equation
(\ref{eq:mm_k1_tpcs}) with $k=0$. Hence $\omega_{\rm s}$ behaves
asymptotically like $\sim |z|^{2\alpha}$. 
The $U(1)$ winding number is
$\nu = \frac{1}{\pi}\int_{\mathbb{C}} \p\bar\p\log
\omega_{\rm s}^{\frac{1}{2}} = \alpha/2$,
which is a half of that of the Abelian case.
The complex parameter $b$ is a realization of the internal 
orientational moduli $\mathbb{C}P^1 \simeq SU(2)/U(1)$ 
which we have explained in Sec.~\ref{sec:u2_cs}. 
Suppose that the Abelian solution has $\hat{\alpha}=2$ (minimal
choice) and hence provides one modulus parameter, then we have found
two moduli parameters in accord with the index result
(\ref{eq:index}), see however Sec.~\ref{sec:uniqueness} for more
comments on this.

\subsubsection{Higher rank gauge group: $U(4)$}

Let us next highlight the vortices in the higher rank gauge group.
For concreteness, we take $G=U(4)$ in this section with $\NF=4$
flavors.
We will explain the vortices not in the full Higgs phase but in the
intermediate and unbroken vacua since the latter ones have not been
studied in literature. 

Let us first consider the following intermediate vacuum
\beq
H = v\ \diag(1,1,0,0) \ ,
\eeq
where $U(4)_{\rm c} \times SU(4)_{\rm f}$ is broken to 
$U(2)_{\rm c} \times S[U(2)_{\rm f} \times U(2)_{\rm c+f}]$.
We can put both the topological and non-topological solitons in this
vacuum. 
For instance, the minimal winding ($k=1$) topological vortex can be
generated by the moduli matrix 
\beq
\Omega = v^{-2}\ \diag(\omega_1,\omega_2,\omega_3,\omega_4) \ , \qquad
H_0 = \diag(z,1,0,0) \ .
\label{eq:ansatz_su4}
\eeq
The corresponding master equations are of the form
\beq
\p\bar\p\log\omega_1 &=& 
\frac{4\pi^2v^4}{\kappa^2}(1-|z|^2\omega_1^{-1})|z|^2\omega_1^{-1} \ ,
\label{eq:mseq_su4_1}\\
\p\bar\p\log\omega_2 &=&
\frac{4\pi^2v^4}{\kappa^2}(1-\omega_2^{-1}) \, \omega_2^{-1} \ ,
\label{eq:mseq_su4_2}\\
\p\bar\p\log\omega_3 &=& 0 \ ,
\label{eq:mseq_su4_3}\\
\p\bar\p\log\omega_4 &=& 0 \ .
\label{eq:mseq_su4_4}
\eeq
The first equation is exactly the same as the master equation in the
Abelian case. 
We solve it with the boundary condition
\beq
\omega_1 \to |z|^2 \ , \qquad 
|z| \to \infty \ .
\eeq
The second one is solved by $\omega_2=1$ and the last two can be
solved as $\omega_{3,4} = {\rm const}$.
We can set the constants to one using a $V$-transformation.
Thus we find
\beq
\Omega = v^{-2}\ \diag(\omega_1,1,1,1) \to 
v^{-2}\ \diag(|z|^2,1,1,1) \ .
\eeq
Let us decompose $\Omega$ into the Abelian and non-Abelian parts
\beq
\Omega = v^{-2} \omega_1^{\frac{1}{4}}\ 
\diag\left(\omega_1^{\frac{3}{4}},
\omega_1^{-\frac{1}{4}},
\omega_1^{-\frac{1}{4}},
\omega_1^{-\frac{1}{4}} \right) \ .
\eeq
From this, we can read off the $U(1)$ winding number and the magnetic
flux 
\beq
\nu &=& \frac{1}{\pi}\int \p\bar\p\log \omega_1^{\frac{1}{4}} =
\frac{1}{4} \ ,\\
\Phi &=& 2\pi \nu \times 4 = 2\pi \ .
\eeq

The moduli matrix in Eq.~(\ref{eq:ansatz_su4}) and the corresponding
master equations (\ref{eq:mseq_su4_1})--(\ref{eq:mseq_su4_4})  
can generate other vortices in the other vacua. 
If we solve Eqs.~(\ref{eq:mseq_su4_1})--(\ref{eq:mseq_su4_4}) with the
boundary condition for $\omega_1$
\beq
\omega_1 \to \frac{|z|^{2+2\alpha}}{C} \ , \qquad 
\text{as}\quad |z| \to \infty \ ,
\eeq
with $\omega_{2,3,4} = 1$, corresponding still to $k=1$, it however
leads to the non-topological soliton in the intermediate vacuum 
\beq
H= v\  \diag (0,1,0,0) \ .
\eeq
Its $U(1)$ winding number and magnetic flux are
\beq
\nu &=& \frac{1}{\pi}\int \p\bar\p\log \omega_1^{\frac{1}{4}} = 
\frac{1+\alpha}{4} \ ,\\
\Phi &=& 2\pi \nu \times 4 = 2\pi (1 + \alpha) \ .
\eeq

Furthermore, if we solve
Eqs.~(\ref{eq:mseq_su4_1})--(\ref{eq:mseq_su4_4}) with the boundary 
condition for $\omega_{1,2}$ 
\beq
\omega_1 \to |z|^2 \ , \quad 
\omega_2 \to \frac{|z|^{2\beta}}{C'} \ , \qquad 
|z| \to \infty \ ,
\eeq
we have the topological vortex with the minimal winding number ($k=1$)
and the non-topological soliton in the different intermediate vacuum 
\beq
H = v\ \diag(1,0,0,0) \ .
\eeq
In this case, $\Omega$ is decomposed as
\beq
\Omega = v^{-2}~\omega_1^{\frac{1}{4}}\omega_2^{\frac{1}{4}}~
\diag\left(
\omega_1^{\frac{3}{4}}\omega_2^{-\frac{1}{4}},~
\omega_1^{-\frac{1}{4}}\omega_2^{\frac{3}{4}},~
\omega_1^{-\frac{1}{4}}\omega_2^{-\frac{1}{4}},~
\omega_1^{-\frac{1}{4}}\omega_2^{-\frac{1}{4}}\right) \ .
\eeq
From this, we can read off the $U(1)$ winding number and the magnetic
flux 
\beq
\nu &=& \frac{1}{\pi}\int \p\bar\p\log \left(\omega_1^{\frac{1}{4}}
\omega_2^{\frac{1}{4}}\right) = \frac{1}{4} + \frac{\beta}{4} \ ,\\
\Phi &=& 2\pi \nu \times 4 = 2\pi + 2\pi\beta \ .
\eeq
Note that, although this form is very similar to the previous example,
the two configurations are completely different. In the previous
example there is a non-topological soliton with a topological soliton
``inside'' and hence flux $2\pi(1+\alpha)$ whereas in the case here,
there are both the topological vortex with the quantized 
magnetic flux $2\pi$ and the non-topological vortex with the flux
$2\pi \beta$.

The last possibility of boundary conditions for
Eqs.~(\ref{eq:mseq_su4_1})--(\ref{eq:mseq_su4_4}) is
\beq
\omega_1 \to \frac{|z|^{2+2\alpha}}{C} \ ,\quad 
\omega_2 \to \frac{|z|^{2\beta}}{C'} \ , \qquad 
|z| \to \infty \ .
\eeq
This leads to the two non-topological solitons in the unbroken vacuum 
\beq
H = \diag(0,0,0,0) \ .
\eeq
The solution has $U(1)$ winding number and magnetic flux as follows 
\beq
\nu &=& \frac{1}{\pi}\int \p\bar\p\log \left(\omega_1^{\frac{1}{4}}
\omega_2^{\frac{1}{4}}\right) = \frac{1+\alpha}{4} + \frac{\beta}{4}
\ ,\\
\Phi &=& 2\pi \nu \times 4 = 2\pi(1+\alpha) + 2\pi \beta \ . 
\eeq

\subsubsection{Higher rank gauge groups: $U(1)\times SO(4)$ and 
$U(1)\times USp(4)$} 

The solitons in the models with $G'=SO(N)$ and $G'=USp(N)$ can also
easily be worked out using the moduli matrix.
To be concrete, let us consider the examples $SO(4)$ and $USp(4)$ in
this section. 
$G'=SO(4),USp(4)$ can be dealt with on the same footing in the moduli
matrix formalism. 
A tiny difference is the sign $\epsilon = \pm 1$ of the invariant
tensor $J$ given in Eq.~(\ref{eq:inv_tensor}).
In what follows, we do not need to distinguish between $G'=SO(4)$ and
$G'=USp(4)$ since all the equations turn out to be the same in these
two cases. 

Let us consider the moduli matrix
\beq
\Omega = v^{-2}~\diag(\omega_1,\omega_1,\omega_2,\omega_2) \ , \quad
H_0 = \diag(z^k,z^k,1,1) \ .
\eeq
With this at hand, the master equation (\ref{eq:masterCS_equal}) is
simplified as 
\beq
\begin{pmatrix}
\p\bar\p\log \omega_1{\bf 1}_2 & \\
& \p\bar\p\log \omega_2{\bf 1}_2
\end{pmatrix}
= \frac{4\pi^2v^4}{\kappa^2}
\begin{pmatrix}
\left(1-|z|^{2k}\omega_1^{-1}\right)|z|^{2k}\omega_1^{-1} {\bf 1}_2 & \\
& \left(1-\omega_2^{-1}\right)\omega_2^{-1} {\bf 1}_2
\end{pmatrix} \ .
\eeq
Note that $\omega_1$ and $\omega_2$ are decoupled since we imposed the
special coupling relation $\kappa = \mu$.

The boundary conditions of $\omega_{1,2}$ determine the type of solitons and
the vacua. If we take the boundary condition
\beq
\omega_1 \to |z|^{2k} \ , \qquad \text{as} \quad 
|z| \to \infty \ ,
\eeq
with $\omega_2=1$, we get $k$ topological (coaxial) vortices in the full
Higgs phase
\beq
H = v~\diag(1,1,1,1) \ .
\eeq
The solution is decomposed as
\beq
\Omega = v^{-2}~\omega_1^{1/2}~
\diag\left(
\omega_1^{1/2}, \omega_1^{1/2}, \omega_1^{-1/2}, \omega_1^{-1/2}
\right) \ .
\eeq
Thus we can read off the $U(1)$ winding number and magnetic flux
\beq
\nu &=& \frac{1}{\pi}\int \p\bar\p\log \omega_1^{1/2} = \frac{k}{2} \ ,\\
\Phi &=& 2\pi \nu \times 4 = 4\pi k \ .
\eeq

If we choose a different boundary condition
\beq
\omega_1 \to \frac{|z|^{2(k+\alpha)}}{C} \ , \qquad \text{as} \quad 
|z| \to \infty \ ,
\eeq
with $\omega_2=1$, it leads to the non-topological soliton in the
intermediate vacuum 
\beq
H = v~\diag(0,0,1,1) \ .
\eeq
The $U(1)$ winding number and the magnetic flux can be read off as
above  
\beq
\nu &=& \frac{1}{\pi}\int \p\bar\p\log \omega_1^{1/2} 
= \frac{k+\alpha}{2} \ ,\\
\Phi &=& 2\pi \nu \times 4 = 4\pi (k+\alpha) \ .
\eeq

If we take a different boundary condition
\beq
\omega_1 \to |z|^{2k} \ , \quad
\omega_2 \to \frac{|z|^{2\beta}}{C'} \ , \qquad \text{as} \quad 
|z| \to \infty \ ,
\eeq
we get $k$ topological (coaxial) vortices as well as the
non-topological vortices in the intermediate vacuum  
\beq
H = v~\diag(1,1,0,0) \ .
\eeq
The solution is decomposed as
\beq
\Omega = v^{-2}~\omega_1^{1/2}\omega_2^{1/2}~
\diag\left(
\omega_1^{1/2}\omega_2^{-1/2}, \omega_1^{1/2}\omega_2^{-1/2}, 
\omega_1^{-1/2}\omega_2^{1/2}, \omega_1^{-1/2}\omega_2^{1/2}
\right) \ .
\eeq
Thus we can read off the $U(1)$ winding number and magnetic flux
\beq
\nu &=& \frac{1}{\pi}\int \p\bar\p\log \left(\omega_1^{1/2}\omega_2^{1/2}\right) 
= \frac{k}{2} + \frac{\beta}{2} \ ,\\
\Phi &=& 2\pi \nu \times 4 = 4\pi k + 4\pi \beta \ .
\eeq

The last possible choice of boundary condition is
\beq
\omega_1 \to \frac{|z|^{2(k+\alpha)}}{C} \ , \quad
\omega_2 \to \frac{|z|^{2\beta}}{C'} \ , \qquad \text{as} \quad 
|z| \to \infty \ .
\eeq
This generates non-topological solitons in the unbroken vacuum
\beq
H = \diag(0,0,0,0) \ .
\eeq
The solution has $U(1)$ winding number and magnetic flux as follows 
\beq
\nu &=& \frac{1}{\pi}\int \p\bar\p\log \left(\omega_1^{1/2}\omega_2^{1/2}\right) 
= \frac{k+\alpha}{2} + \frac{\beta}{2} \ ,\\
\Phi &=& 2\pi \nu \times 4= 4\pi (k+\alpha) + 4\pi \beta \ .
\eeq

As in the case of $U(4)$ CS theory, the non-topological
solitons in $U(1)\times SO(4)$ and $U(1)\times USp(4)$ carry
non-Abelian zero-modes associated with the spontaneous symmetry
breaking. 
To see this, let us consider a minimal example with the embedding solution 
\beq
H = \diag(H^{\rm NTP},H^{\rm NTP},0,0) \quad \to \quad
\left\{
\begin{array}{lcl}
{\rm diag}(v',v',0,0) & & \text{at the origin}\\
{\rm diag}(0,0,0,0) & & \text{at the infinity}
\end{array}
\right..
\eeq
The global symmetry respected at spatial infinity is $SU(4)_{\rm f}$
which however is spontaneously broken down to 
$U(2)_{\rm c+f} \times SU(2)_{\rm f}$ by the soliton (the same is true
for $USp(4)$). 
Thus the non-Abelian moduli space  is given by the Grassmannian 
\beq
{\cal M}_{\rm orientation} = \frac{SU(4)}{U(2)\times SU(2)} \simeq
Gr_{4,2} \ .
\eeq
Note that this is different from the orientational moduli $SO(4)/U(2)$
($USp(4)/U(2)$) of the minimal topological soliton in the $SO(4)$
($USp(4)$) model. 
We have observed that the topological and non-topological solitons
accidentally have the same orientational moduli
$SU(N)/U(N-1) \simeq \mathbb{C}P^{N-1}$ in the $U(N)$ CS
model. This is however in general not the case. 

\section{On the uniqueness of the master
  equations \label{sec:uniqueness}} 

In this section, we will try to solve the uniqueness problem of
the master equations (\ref{eq:YMmasterA})--(\ref{eq:YMmasterNA}) or 
(\ref{eq:CSmasterA})--(\ref{eq:CSmasterNA}), 
which we pointed out in Sec.~\ref{sec:modulimatrix}.
For that purpose, we will consider small fluctuations around the gauge 
invariant field $\Omega$
\beq
\Omega \to \Omega_{\rm s} + \delta \Omega,
\label{eq:fluc}
\eeq
with $\Omega_{\rm s}=S_{\rm s}S_{\rm s}^\dag$ being a true solution
corresponding to a given configuration $\Omega_0$. 
If the master equation has the uniqueness property\footnote{
Since, at the boundary $|z|\to\infty$, the solution $\Omega_s$ is
known in terms of the lump solution for the semi-local topological solitons
(which is the statement that
$\Omega_s$ is in the vacuum manifold at $|z|\to\infty$) and
furthermore that this solution is an algebraic solution and unique, we
know that even if there exist another solution $\Omega$ it has to obey
the \emph{same} boundary condition at $|z|\to\infty$ and hence it
follows that 
$\delta\Omega \to 0$ for $|z|\to\infty$.
}, 
$\delta\Omega$
must be zero in the whole $\mathbb{C}$-plane. 
So our goal is to confirm that $\delta\Omega = 0$.


\subsection{Yang-Mills-Higgs theory}

In order not to introduce unessential complication to the following
argument, let us consider the equal gauge coupling case $e=g$. The
master equation in focus is Eq.~(\ref{eq:masterYM_equal}).
The fluctuations in Eq.~(\ref{eq:fluc}) obey to the linear order
\beq
\bar{\p}\p\phi
+\left[\Omega_{\rm s}^{-1}\bar{\p}\Omega_{\rm s},\p\phi\right]
= \frac{e^2}{2} \Omega_{\rm s}^{-1} 
\left<\Omega_0\,\phi\,\Omega_{\rm s}^{-1}\right>_G\Omega_{\rm s} \ , 
\eeq
where we have defined the $N$-by-$N$ gauge invariant matrix field 
\beq
\phi \equiv \Omega_{\rm s}^{-1}\delta \Omega \ .
\eeq
Note that since $\phi^\dag$ and $\bar\p\phi^\dag$ transform as
holomorphic adjoint fields with respect to the $V$-transformation 
\beq
\phi^\dag \to V \phi^\dag V^{-1} \ , \qquad
\bar\p \phi^\dag \to V \bar\p \phi^\dag V^{-1} \ .
\eeq
then the above equation can be further rewritten as
\beq
\left(\D_v \bar\p \phi^\dag \right)^\dag
= \frac{e^2}{2} 
\left<\Omega_{\rm s}^{-1}\Omega_0\,\phi\right>_G \ , 
\label{eq:fluc_mas_YM} 
\eeq
where we have used the holomorphic covariant derivative
(\ref{eq:coDv}) and 
\beq
\Omega_{\rm s}^{-1} 
\left< X \Omega_{\rm s}^{-1}\right>_G 
\Omega_{\rm s}
= \Tr\left[X \Omega_{\rm s}^{-1}t^\alpha\right]
\Omega_{\rm s}^{-1} t^\alpha \Omega_{\rm s}
= \Tr\left[\Omega_{\rm s}^{-1} X  \hat{t}^\alpha\right]
\hat{t}^\alpha
= \left<\Omega_{\rm s}^{-1} X \right>_G \ ,
\label{eq:relation_omega}
\eeq
with $\hat{t}^\alpha \equiv \Omega_{\rm s}^{-1}t^\alpha\Omega_{\rm s}$.

We compare Eq.~(\ref{eq:fluc_mas_YM}) with
Eq.~(\ref{eq:schrodinger_YM}) by changing the variable from the adjoint
field $\chi_{\rm a}$ to a gauge invariant field given by 
\beq
\psi^\dag \equiv S_{\rm s}\chi_{\rm a} S_{\rm s}^{-1} \ .
\label{eq:psi_def}
\eeq
With respect to $\psi$, Eq.~(\ref{eq:schrodinger_YM}) is expressed as
\beq
\left(\D_v \bar\p \psi^\dag \right)^\dag
= \frac{e^2}{2} \Tr\left[\Omega^{-1}\Omega_0\psi \tilde{t}^\alpha\right] \tilde{t}^\alpha
= \frac{e^2}{2} \left<\Omega^{-1}\Omega_0\psi\right>_G,
\label{eq:fluc_mas_YM2}
\eeq
with $\tilde{t}^\alpha\equiv(S_{\rm s}^{\dag})^{-1}t^\alpha S_{\rm s}^\dag$.
Thus, the fluctuation $\phi$ of the master equation and the zero-mode
$\psi$ of the adjoint operator $\mathbb{D}^\dag$ obey exactly the same
equations (\ref{eq:fluc_mas_YM}) and (\ref{eq:fluc_mas_YM2}). 

Since we have already proven that $\psi=0$ ($\chi_{\rm a}=0$) in
Sec.~\ref{sec:vanish_YM}, we can immediately conclude that 
$\phi = 0$. Thus we have proven the uniqueness of the master equation.
Then we conclude that all the zero-modes reside 
in the moduli matrix $H_0$ and $\Omega$ has no moduli parameters.

Note that this uniqueness is {\it local} uniqueness since $\phi$ is only a
small fluctuation around the true solutions. 
We still do not provide any proof for the global uniqueness. This is
one of the important future problems. See however the discussion in
Sec.~\ref{sec:conclusion}.


\subsection{Chern-Simons-Higgs theory \label{sec:unique_cs}}

As in the YM case, we take equal couplings $\kappa=\mu$ and
consider a small fluctuation around the moduli matrix field 
$\Omega\to \Omega_{\rm s} + \delta\Omega$ which by plugging into the
master equation (\ref{eq:masterCS_equal}) yields to linear order 
\begin{align}
\left(\D_v \bar\p \phi^\dag \right)^\dag
&=\frac{16\pi^2}{\kappa^2}
\Omega_{\rm s}^{-1} \left(
\left<\Omega_0\Omega_{\rm s}^{-1}
  \left<\Omega_0\phi\Omega_{\rm s}^{-1}\right>_G\right>_G
+\left<\Omega_0\phi\Omega_{\rm s}^{-1}
  \left<\Omega_0\Omega_{\rm s}^{-1}-v^2\mathbf{1}_N\right>_G\right>_G
\right)\Omega_{\rm s} \ ,
  \nonumber
\end{align}
where we have used the identity (\ref{eq:bracketidentity}) before
inserting the fluctuations and we have again used the gauge invariant
field $\phi\equiv\Omega_{\rm s}^{-1}\delta\Omega$. As in the
YM case we can use (\ref{eq:relation_omega}) to rewrite the
above equation as 
\begin{align}
\left(\D_v \bar\p \phi^\dag \right)^\dag
&=\frac{16\pi^2}{\kappa^2}
\left(
\left<\Omega_{\rm s}^{-1}\Omega_0
  \left<\Omega_{\rm s}^{-1}\Omega_0\phi\right>_G\right>_G
+\left<\Omega_{\rm s}^{-1}\Omega_0\phi
  \left<\Omega_{\rm s}^{-1}\Omega_0-v^2\mathbf{1}_N\right>_G\right>_G
\right) \ .
\label{eq:CSmasterfluc}
\end{align}

Next, we would like to compare Eq.~(\ref{eq:CSmasterfluc}) with
Eq.~(\ref{eq:CSvanishingeqcombined}) which the zero-modes 
$\chi_{\rm a}$ of $\mathbb{D}^\dag$ in Eq.~(\ref{eq:adj_cs}) obey,
as in the case of the YM system.
Using again the change of variables (\ref{eq:psi_def}), we can write 
Eq.~(\ref{eq:CSvanishingeqcombined}) on the form 
\begin{align}
\left(\D_v \bar\p \psi^\dag \right)^\dag
&=\frac{16\pi^2}{\kappa^2}\left(
\left<\Omega_{\rm s}^{-1}\Omega_0
  \left<\Omega_{\rm s}^{-1}\Omega_0\psi\right>_G\right>_G
+\left<\Omega_{\rm s}^{-1}\Omega_0\psi
  \left<\Omega_{\rm s}^{-1}\Omega_0 -
  v^2\mathbf{1}_N\right>_G\right>_G \right) \ ,
\label{eq:CSmasterfluc2}
\end{align}
where
$\left<X\right>_G=\Tr\left[X\tilde{t}^\alpha\right]\tilde{t}^\alpha$
with $\tilde{t}^\alpha\equiv (S^\dag)^{-1}t^\alpha S^\dag$.
Hence, the fluctuation $\phi$ of the master equation and the zero-mode 
$\psi$ of the adjoint operator $\mathbb{D}^\dag$ obey exactly the same 
equations, namely Eqs.~(\ref{eq:CSmasterfluc}) and (\ref{eq:CSmasterfluc2}).

Although $\psi$ and $\phi$ obey exactly the same equation, they are 
different fields. While $\phi$ is a small fluctuation around 
$\Omega_{\rm s}=S_{\rm s}S_{\rm s}^\dag$ which is related to the
physical field configurations $H$ and $A_i$ through Eq.~(\ref{eq:mm}),
$\psi$ contains the normalizable zero-modes of $\mathbb{D}^\dag$ and
has nothing to do with $H$ and $A_i$. This difference must be
carefully dealt with, because $\phi$ is not necessarily a bounded
solution especially in the case of the non-topological solitons. 
What we should require the normalizability of, is not the small
fluctuations of $\Omega$ but of the physical fields $A_i$ and $H$,
namely $\delta\!A_i$ and $\delta\!H$ must be normalizable.

To be concrete, let us first try to find additional moduli parameters 
of non-topological solitons in the Abelian CS case
\cite{Lee:1991td} by taking advantage of the moduli matrix
method\footnote{Here we do not fix any gauge unlike
  Ref.~\cite{Lee:1991td} where the authors worked in the Coulomb
  gauge. Instead, we deal with the gauge invariant quantity, namely
  $\omega$ and its fluctuation.}. 
Given a moduli matrix $H_0(z)=1$, we consider a fluctuation 
$\phi = \omega^{-1}\delta\omega$ which asymptotically satisfies
\beq
\p\bar\p\phi = 
\frac{4\pi^2}{\kappa^2}\left(2\omega^{-1}-v^2\right)\omega^{-1} \phi
\simeq - \frac{4\pi^2 v^2}{\kappa^2} \frac{C}{|z|^{2\alpha}} \phi \ ,
\label{eq:fluc_cs_asym}
\eeq
where we have used $\omega \simeq |z|^{2\alpha}/C$.
Furthermore, the variation in the magnetic flux density is given by
\beq
F_{12} + \delta F_{12} =  -2 \bar\p\p \log \omega -2 \bar\p\p \phi
\to -2 \bar\p\p \log \frac{|z|^{2\alpha}}{C} -2 \bar\p\p \phi \ ,
\qquad {\rm for} \ |z| \to \infty \ . 
\eeq 
In order not to change the total energy of the soliton, the second
term must vanish. 
Therefore, the fluctuation $\phi$ will asymptotically approach the
real part of the holomorphic function as
\beq
\phi \to F_0(z) + F_0^*(\bar z) \ , \qquad
F_0(z) = \sum_{i=1}^f a_iz^i \ .
\eeq
From Eq.~(\ref{eq:fluc_cs_asym}), $f \in \mathbb{N}_{\ge 1}$ should be 
$f < 2\alpha$.
Note that we have suppressed the constant term ($i=0$). This is
because the asymptotic behavior of $\omega$ is fixed as 
$\omega \to |z|^{2\alpha}/C$. 
For instance, if we consider $F_0= a_0$ ($a_0 \in \mathbb{C}$), we
have 
$\omega' = \omega + \delta\omega \to \omega(1+2 {\rm Re}[a_0]) 
\simeq \frac{1+2 {\rm Re}[a_0]}{C}\, |z|^{2\alpha}$.
Since we have chosen, however, the asymptotic behavior 
$\omega \to |z|^{2\alpha}/C$ in such a way that the configuration
becomes regular, we have to choose $a_0 = 0$. 
Finally, we should impose the normalizability condition on
$\delta\!H$ and $\delta\!\bar{A}$.
Fluctuations of $H$ and $\bar{A}$ are given by
\beq
\delta\!H = - F H \ ,\qquad
\delta\!\bar{A} = -i \bar{\p} F \ , \qquad 
F \equiv s^{-1}\delta s \ .
\eeq
$\phi$ and $F$ are related by $\phi = F + F^*$. Thus we can
asymptotically identify $F \to F_0(z)$ and in turn 
$\delta\!\bar{A} \to 0$ is automatically ensured.
Moreover, normalizability of $\delta\!H$ which behaves as 
$\delta\!H \to -F_0 |z|^{-\alpha}$ requires that the power of $F_0$
should be $f=\hat{\alpha}-1$. 
Hence, we obtain the boundary condition of $\phi$ and $\omega$
\beq
\phi &\to& {\rm Re}\left[\sum_{i=1}^{\hat\alpha -1} a_i z^i\right] ,\\
\omega &\to& C^{-1}|z|^{2\alpha}\left(1 + \sum_{i=1}^{\hat\alpha -1} a_i z^i + {\rm c.c.} \right).
\label{eq:bc_cs}
\eeq
This supplies $\hat{\alpha}-1$ additional complex parameters 
$\{a_i\}$ ($i=1,2,\ldots,\hat{\alpha}-1$) in $\omega$ as varieties of the boundary 
condition for $\omega$.
Note that this result is not consistent with the index theorem result
${\cal I} = \hat \alpha$ ($\NF=N=n_0=1$) given in Eq.~(\ref{eq:index_cs_nt}).
This mismatch was first observed in Ref.~\cite{Jackiw:1990pr}.
In this way, the index sometimes over-counts the zero-modes. So we
should be careful when we count the physical zero-modes using the
index theorem, i.e.~we should correctly subtract unphysical zero-modes. 
This problem seems however only to arise in the non-topological cases. 

Therefore, we conclude that $\omega$ has its own zero-modes in addition to
those in the moduli matrix for the non-topological solitons.

This is in sharp contrast to the case of the topological CS solitons 
in which there are no degrees of freedom in $\omega$ 
once the moduli matrix is given.
This can be seen as follows. The fluctuation $\phi$ asymptotically
satisfies 
\beq
\bar{\p}\p \phi \simeq m_{\kappa\infty}^2 \phi \ ,
\eeq
where we have used the asymptotic form of $\omega\sim |z|^{2k}/v^2$. 
Clearly, the right-hand side does not allow $\phi$ to be a harmonic 
function and there is no normalizable solution compatible with the
fact that $\phi$ needs to be the real part of a holomorphic polynomial
and hence we can conclude that $\phi$ must vanish.

Once we fix the boundary condition of $\omega$, namely we set $\phi \to 0$,
we can utilize the similarity between $\phi$ and $\psi$ which both 
satisfy the boundary condition $\phi,\psi \to 0$ at infinity. 
Since we have already shown that plausibly 
$\psi=0$ ($\chi_{\rm a}=0$) in Sec.~\ref{sec:vanish_CS}, we expect
in general that $\phi = 0$, so that the master equation is unique.
Note that this uniqueness is local uniqueness since $\phi$ is only a
small fluctuation around the true solutions. As in the case of
YM solitons, the global uniqueness problem is an important
future problem.

Let us next consider the non-topological non-Abelian CS
solitons. 
As in the Abelian case, we consider a variation in the magnetic flux
density  
\beq
\delta \Tr[F_{12}] 
= -2\delta \left(\p\bar\p\Tr\log \Omega\right) 
= -2 \p\bar\p\,\Tr[\phi] \ ,
\eeq
where $\phi = \Omega^{-1}_{\rm s}\delta\Omega$ is an $N$-by-$N$ matrix
field. 
The contribution to the total magnetic flux from $\phi$ must be zero,
so we require the following boundary condition for $\phi$:
\beq
\phi \to \Omega_{\rm s}^{-1} F_0(z) \Omega_{\rm s} 
 + F_0^\dag (\bar{z}) \ ,
\label{eq:phi_F}
\eeq
where $F_0(z)$ is an arbitrary $N$-by-$N$ holomorphic matrix.
With this boundary condition, we have to solve
Eq.~(\ref{eq:CSmasterfluc}). 
This holomorphic matrix $F_0(z)$ supplies additional zero-modes
to the non-topological solitons.
In order to find a condition on $F_0(z)$, let us consider the
fluctuation of $\bar{A}$ and $H$ which can be written as
\beq
\delta\!H = - S^{-1} F S H \ , \qquad
\delta\!\bar{A} = -i S^{-1}\bar\p F S \ , \qquad
F \equiv \delta S S^{-1} \ .  
\eeq
Note that $F$ is related to $\phi$ by 
\beq
\phi = \Omega^{-1}_{\rm s} F \Omega_{\rm s} + F^\dag \ .
\eeq
Comparing this with Eq.~(\ref{eq:phi_F}), we can identify $F_0(z)$ as
the asymptotic function of $F$, namely 
$F(z,\bar z) \to F_0(z)$ as $|z|\to\infty$.
The fluctuations of the physical fields $H$ and $\bar{A}$ must 
asymptotically go to zero. 
The holomorphy of $F_0(z)$ is indeed needed for 
$\delta\!\bar{A}$ to vanish asymptotically. 
Furthermore, powers of the holomorphic functions in $F_0(z)$ are
determined by the normalizability condition on $\delta\!H$. 

Let us give some examples in order to understand better the
situation. 
For simplicity, let us take $G = U(2)$ and consider the
non-topological soliton in the unbroken vacuum $H=\diag(0,0)$. To
create the vortex, we take the moduli matrix and make the diagonal
Ansatz for $\Omega_{\rm s}$ 
\beq
H_0 = \diag(1,0) \ , \qquad
\Omega_{\rm s} = \diag(\omega_1,1) \ .
\eeq
We choose the asymptotic behavior 
$\omega_1 \sim |z|^{2\alpha}/C$ as $|z| \to \infty$.
Note that this moduli matrix does not completely fix the
$V$-equivalence. 
The residual infinitesimal $V$-transformation takes the form
\beq
V = \begin{pmatrix}
1 & \delta_1(z)\\
0 & 1+\delta_2(z)
\end{pmatrix}.
\eeq
This transforms $\Omega_s$ as
\beq
\delta \Omega_{\rm s}  = 
\begin{pmatrix}
0 & \delta_1\\
\delta_1^* & \delta_2 + \delta_2^*
\end{pmatrix} \ .
\label{eq:v_resi}
\eeq
Now we are ready to look at the fluctuations at infinity.
To this end, we write $F_0(z)$ as
\beq
F_0 = \begin{pmatrix}
f_{11}(z) & f_{12}(z)\\
f_{21}(z) & f_{22}(z)
\end{pmatrix} \ .
\label{eq:F0_def}
\eeq
Then the fluctuations are expressed by
\beq
\delta \Omega &=& F \Omega_{\rm s} + \Omega_{\rm s} F^\dag \to
\begin{pmatrix}
\omega_1(f_{11} + f_{11}^*) & \omega_1 f_{21}^* + f_{12}\\
\omega_1 f_{21} + f_{12}^* & f_{22} + f_{22}^*
\end{pmatrix} \ , \\
\delta\! H &=& - S^{-1} F H_0 \to 
-
\begin{pmatrix}
\omega_1^{-1/2} f_{11} & 0\\
f_{21} & 0
\end{pmatrix} \ .
\eeq
We impose the square integrability condition on the fluctuations
\beq
\Tr\left[\delta\! H \delta\! H^\dag\right]
\to \Tr\left[ \Omega_{\rm s}^{-1}F_0 \Omega_0^\dag F_0^\dag \right]
= \omega_1^{-1}|f_{11}|^2 + |f_{21}|^2  \ .
\eeq
Hence, we find that the most generic form of the fluctuations is given
as 
\beq
f_{11} = \sum_{i=1}^{\hat\alpha -1} a_iz^i \ , \quad 
f_{21} = 0 \ .
\eeq
On the other hand, there are no constraints for $f_{12}$ and $f_{22}$,
it seems that $f_{12}$ and $f_{22}$ have an infinite number of
additional zero-modes. 
However, they are unphysical because they can be eliminated by the
residual $V$-transformation given in Eq.~(\ref{eq:v_resi}). 
Hence, we have found $\hat \alpha - 1$ additional moduli parameters in
$f_{11}(z)$. 
Furthermore, we have found the orientational moduli parameter $b$
given in Eq.~(\ref{eq:mm_cs_nt_ori}).
So in total we have found 
$\hat\alpha - 1 + 1 = \hat \alpha$ complex parameters for the 
non-topological solitons.
Again, the result is not consistent with the index theorem result
${\cal I} = 2 \hat\alpha$ given in Eq.~(\ref{eq:index_cs_nt}).
At this stage, we are not certain if the index over-counts the number
of physical zero-modes $\hat\alpha$ by including unphysical modes or
we simply did not exhaust all the possible zero-modes in the above
moduli matrix calculation. We will leave this problem for a future
work. 

Let us consider another configuration by taking the following moduli
matrix 
\beq
H_0 = \diag(1,1) \ , \qquad
\Omega_{\rm s} = 
\diag (\omega_1, \omega_2) \ .
\eeq
We fix the boundary conditions as $\omega_1 \to |z|^{2\alpha_1}/C_1$
and $\omega_2 \to |z|^{2\alpha_2}/C_2$, while we define 
$\alpha \equiv \alpha_1 + \alpha_2$. 
Note that there are no residual $V$-equivalence relations in this
case. The fluctuations are written as (where $F_0(z)$ still is given
by Eq.~(\ref{eq:F0_def}))
\beq
\delta \Omega &=& F \Omega_{\rm s} + \Omega_{\rm s} F^\dag \to
\begin{pmatrix}
\omega_1(f_{11} + f_{11}^*) & \omega_1 f_{21}^* + \omega_2 f_{12}\\
\omega_1 f_{21} + \omega_2 f_{12}^* & \omega_2\left(f_{22} + f_{22}^*\right)
\end{pmatrix} \ , \\
\delta\! H &=& - S^{-1} F H_0 \to 
-
\begin{pmatrix}
\omega_1^{-1/2} f_{11} & \omega_1^{-1/2} f_{12}\\
\omega_2^{-1/2} f_{21} & \omega_2^{-1/2} f_{22}
\end{pmatrix} \ .
\eeq
As before, we impose the square integrability condition
\beq
\Tr\left[\delta\! H \delta\! H^\dag\right]
\to \Tr\left[ \Omega_{\rm s}^{-1}F_0 \Omega_0^\dag F_0^\dag \right]
= \omega_1^{-1}\left(|f_{11}|^2 + |f_{12}|^2\right) + 
 \omega_2^{-1} \left(|f_{21}|^2+|f_{22}|^2\right) \ .
\eeq
This gives the upper bounds for $f_{IJ}$ ($I,J=1,2$)
\beq
f_{II} &=& \sum_{i=1}^{\hat\alpha_I-1} a_i^{(II)}z^i \ , \\
f_{IJ} &=& \sum_{i=0}^{\hat\alpha_I-1} a_i^{(IJ)}z^i , \qquad 
(I\neq J) \ .
\eeq
Here we did not count $i=0$ for $f_{II}$ because the constant part is 
fixed by the boundary condition $\omega_I \to |z|^{2\alpha_I}/C_I$.
Thus we found 
$(\hat\alpha_1 - 1) + \hat\alpha_1 +  (\hat\alpha_2 - 1) + \hat\alpha_2 
= 2 (\hat\alpha_1 + \hat\alpha_2 - 1)$ 
moduli parameters in $\{f_{IJ}\}$.
Again, there is a mismatch with the index theorem result
(\ref{eq:index_cs_nt}), see the comments above.

\section{Conclusion and discussion\label{sec:conclusion}}

In this paper we have studied the non-Abelian solitons in the
${\cal N}=2$ supersymmetric gauge theories in three dimensions;
viz.~the YM and CS gauge theories. 
We have used a common framework for investigation of the non-Abelian
solitons which nicely works both for the YM theories and CS theories.
We have found several new solitons; the non-Abelian non-topological
solitons and the non-Abelian semi-local (topological/non-topological)
solitons in the unbroken and the partially broken vacua which exist in
the CS theories. 
The non-Abelian non-topological solitons have internal orientational 
moduli analogous to the non-Abelian topological solitons, which
however is in general different from that of the topological solitons. 
Furthermore, to the best of our knowledge, the present paper is the
first work which has found the semi-local solitons in the models with
the number of flavors less than or equal to the number of colors. 

In addition to the minimal solution, we have also paid great attention
to the multiple solitons and their zero-modes.
To this end, we have made use of two methods, counting the number of 
zero-modes by the index theorem and realizing the explicit zero-modes
by the moduli matrix formalism. 
Those for the YM solitons given in Sec.~\ref{sec:index_YM} and
\ref{sec:mm_ym} are all known and we have simply reviewed them.
What we learned from these review parts of the YM vortices is
that the moduli matrix formalism is a rigorous tool for investigating
topological solitons\footnote{Needless to say, the moduli matrix
  formalism has been already established. Many results obtained using
  the moduli matrix formalism are summarized in the review
  \cite{Eto:2006pg}.}. 
Indeed, the index theorem counting \cite{Hanany:2003hp}
and the number of the zero-modes realized in the moduli matrix
formalism \cite{Eto:2005yh} do in fact coincide.
In order to comfirm that the moduli matrix formalism indeed provides
us with all possible moduli parameters, we have to solve the
long-standing uniqueness and existence problems of the master
equations (\ref{eq:YMmasterA}) and (\ref{eq:YMmasterNA}). 
As to the uniqueness problem, we have made a progress. We have proven
the local uniqueness of the master equation by finding a one-to-one
correspondence between the small fluctuations of the master equations 
and the vanishing theorem.

We have also tried to count the zero-modes by the index theorem for
the CS theories, although we could not prove the vanishing theorem. 
Furthermore, in order to go beyond the parameter counting, we have
also applied the moduli matrix formalism which has already been used
to investigate the topological solitons in
Refs.~\cite{Gudnason:2009ut,Gudnason:2010yy}. For the topological
solitons, we have reached at the same level as the YM case, namely the 
coincidence of the index theorem result and the number of the
zero-modes realized in the moduli matrix. 
Actually, we have found no differences both in the index theorem
calcuation and the moduli matrix formalism for the topological
solitons between the YM and CS theories, except for some technical 
details. 
This fact strongly suggests that the moduli matrix formalism is also
a rigorous tool for investigating the CS solitons. If this is the
case, the moduli space of the topological CS solitons is completely
the same as that of the YM theory which has been studied intensively, 
which was conjectured in \cite{Gudnason:2009ut}. 
Even the metric of the moduli spaces are coincident in the two
theories (the dynamics is however different due to the CS term
\cite{Collie:2008mx}), so there is nothing new about the moduli space
of the topologial solitons. 

However, this is not the end of story. Proper features of the CS
theory appear when we study the non-topological and semi-local
solitons in the unbroken and the partially broken vacua. 
We have also made use of the moduli matrix formalism and showed it
works well too. 
A remarkable contrast to the YM theory is that the gauge invariant
field $\Omega$ has its own moduli parameters in addition those
provided by the moduli matrix, namely, there are zero-modes in the  
master equation. 
We have found them in the varieties of the harmonic boundary condition
for the master equation. Once $H_0$ and the boundary condition of
$\Omega$ are properly fixed, we would expect in general that the
master equation in the CS theory is unique at least locally. 
We have given several examples, 
$G=U(2),U(4),U(1)\times SO(4)$ and $U(1)\times USp(4)$ where the
Abelian solitons can be embedded solutions in the special case of
equal CS couplings. 
The generalization to the higher rank gauge groups is
straightforward.

We have attacked the long-standing problem of the uniqueness
properties of the master equations
and we have proved completely independent of the gauge group that a
small fluctuation of the gauge invariant field $\Omega$ 
around the topological solitons is in one-to-one correspondence with  
the vanishing theorem of the index theorem calculation. 
\begin{figure}[!t]
\begin{center}
\includegraphics[width=0.4\linewidth]{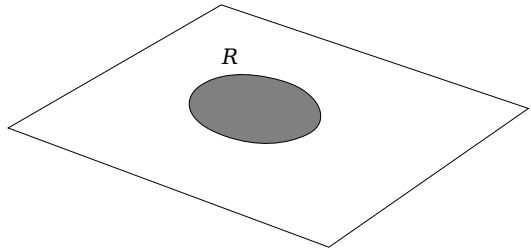}
\caption{{\small The fluctuations of the master equation outside the
    radius $R$ have to be infinitesimal, since the fluctuations have
    to vanish identically at infinity. Inside $R$ they could become
    finite. However, we have shown in Sec.~\ref{sec:uniqueness} that
    the infinitesimal variation has to vanish. }}
\label{fig:radius}
\end{center}
\end{figure}

There is one subtlety, but as we shall argue, physically not too
worrying fact about this uniqueness calculation. In this discussion
here, we will consider only the question of local uniqueness proved
for the YM case and expected for the topological CS case.
The reason for which we only claim that the uniqueness is local is
that the fluctuations of the master equation considered are just
infinitesimal fluctuations. 
Hence, one would immediately ask, what if the fluctuation is finite?
Is there then a possibility for the existence of another solution and
hence a parameter governing the different solutions? 
Let us first recall that in the vacuum, the solution to the master
equation is completely unique in all cases, i.e.~it is the well-known
lump solution (even if it is technically singular in the local vortex
case). Hence at an infinite radius, the fluctuations must vanish  
identically. Now let us assume that there exist two solutions,
$\Omega_{A,B}$. The finite difference is necessarily zero at
infinity. Consider now a finite radius $R$ (see Fig.~\ref{fig:radius})
much larger than the typical scale of the vortex system $\mu$. Let us
assume that the finite difference could be non-zero within the
radius $R$. However, from that radius till infinity the difference is 
expected to be very small, say of the order of $1/(\mu R) \ll 1$. Thus
we expect the difference to be exactly an infinitesimal deviation from
the true solution. However, our uniqueness calculation shows that if
the vanishing theorem is satisfied, then there cannot be a non-zero 
fluctuation deviating from a true solution. 
The somewhat physical argument tells us that in this case, we can
expect the difference $\Omega_B - \Omega_A$ to vanish. 
A more rigid demonstration of this difficult problem is indeed
welcome and left for future works.

We will now discuss what we were not able to do in this work. 
For the YM theory, the global uniqueness and the existence problems of 
the master equation are unsolved\footnote{Clearly, the existence of
  the embedded Abelian solution exists provided the gauge couplings
  are equal. What we mean here is that the solution exists in general
  for all values of the couplings and moduli parameters.}. 
For the CS theory, 1) the proof of the vanishing theorem,
2) the uniqueness and exisitence problems, 
3) matching between the parameter counting of the index theorem and
the moduli matrix formalism for the non-topological and semi-local
solitons, 
4) generic solutions of the non-topological and semi-local solitons,
especially with the intermediate relative orientational moduli 
and 
5) numerical solutions for the non-topological and semi-local vortices
for generic coupling $\kappa \neq \mu$. 
We left these problems for future works.

A comment in store is about the equal gauge coupling choice which we
have made in this paper. It causes no problem for the number of moduli
in a solution. Our derivations of all the generic formulae done for
equal gauge couplings can trivially be extended to the generic gauge
coupling case. We have made this choice for simplicity and aesthetic
beauty.  
However, the solution to a given master equation (CS or
YM case) is clearly different in the case of different gauge
couplings. Hence, for the realization of explicit solutions, it is
certainly important to consider the generic gauge coupling case and
verify the existence of the solutions for all (finite) values of the
couplings and values of the moduli parameters. 

Another interesting thing might be the realization of the low-energy
effective theory on the non-topological CS solitons. For the
topological CS solitons it has been studied in 
Refs.~\cite{Aldrovandi:2007nb,Collie:2008mx,Collie:2008za}. 
Dynamics and interactions of non-Abelian solitons, especially
non-topological solitons, which may depend on the orientations are
also interesting open problems. 

Non-Abelian solitons in the CS theory with the YM kinetic term
may have similar properties compared with the solitons in the current
work. 
The D-brane construction of the moduli space is another interesting
problem. Furthermore, the non-relativistic limit may be also 
interesting, especially due to the integrability found in the Abelian
non-relativistic systems \cite{Jackiw:1990tz}.

\subsubsection*{Acknowledgments}

The authors would like to thank Jarah Evslin, Koji Hashimoto, Kenichi
Konishi, Akitsugu Miwa and Keisuke Ohashi for useful comments and
discussions. 
The work of M.E. is supported by Special Postdoctoral Researchers
Program at RIKEN. 


\end{document}